\documentclass[twocolumn]{aastex62}
\usepackage{amssymb}
\usepackage{amsmath}
\usepackage[]{graphicx}
\usepackage{enumerate}
\usepackage{mathrsfs,amssymb}

\usepackage{color} 
\usepackage{xspace}

\tightenlines

\shorttitle{Outflows with cosmic rays and waves}
\shortauthors{Ramzan, Ko \& Chernyshov}

\begin{document}

\title{Outflows in the presence of cosmic rays and waves}

\author{B. Ramzan}
\affiliation{Institute of Astronomy, National Central University, Zhongli Dist., Taoyuan City, Taiwan (R.O.C.)}
\author{C. M. Ko}
\affiliation{Institute of Astronomy, National Central University, Zhongli Dist., Taoyuan City, Taiwan (R.O.C.)}
\affiliation{Department of Physics and Center for Complex Systems, National
Central University, Zhongli Dist., Taoyuan City, Taiwan (R.O.C.)}
\author{D. O. Chernyshov}
\affiliation{I. E. Tamm Theoretical Physics Division of P. N. Lebedev Institute of Physics, Leninskii pr. 53, 119991 Moscow,
Russia}
\affiliation{Moscow Institute of Physics and Technology (State University), 9, Institutsky lane, Dolgoprudny, 141707, Russia}

\correspondingauthor{C. M. Ko}
\email{cmko@astro.ncu.edu.tw}


\begin{abstract}
Plasma outflow or wind against a gravitational potential under the influence of cosmic rays is studied in the context of hydrodynamics.
Cosmic rays interact with the plasma via hydromagnetic fluctuations.
In the process, cosmic rays advect and diffuse through the plasma.
We adopt a multi-fluid model in which besides thermal plasma, cosmic rays and self-excited Alfv\'en waves are also treated as fluids.
We seek possible physically allowable steady state solutions of three-fluid (one Alfv\'en wave) and four-fluid (two Alfv\'en waves) models
with given the boundary conditions at the base of the potential well.
Generally speaking, there are two classes of outflows, subsonic and supersonic (with respect to a suitably defined sound speed).
Three-fluid model without cosmic ray diffusion can be studied in the same way as the classic stellar wind problem, and is taken as a reference model.
When cosmic ray diffusion is included, there are two categories of solutions.
One of them resembles the three-fluid model without diffusion,
and the other behaves like thermal wind at large distances when the waves wither and cosmic rays are decoupled from the plasma.
We also inspect the effect of wave damping mechanisms
(such as, nonlinear Landau damping).
Roughly speaking, the effect is much smaller in supersonic outflow than in subsonic outflow.
\end{abstract}


\section{Introduction}
\label{sec:introduction}

High energy activities, such as supernovae, energetic winds, compact objects, produce energetic charged particles (a.k.a. cosmic rays)
in our Galaxy and others. Although the energy per nucleon of the highest energy cosmic ray particles can be up to $10^{20}$ eV,
the energy density of cosmic ray in our Galaxy is dominated by low energy cosmic rays (of the order of hundreds of MeV).
Interestingly enough, in our Galaxy cosmic ray energy density and the energy density of other components of the interstellar medium
(e.g., different phases of gas, magnetic field) are of the same order of magnitude \citep[e.g.,][]{Parker_1969,Ginzburg_1976,Ferrire_2001,Cox_2005}.
The seemingly equipartition of energy indicates there are significant interactions among the components of the interstellar medium.
Thus, cosmic rays is expected to play a dynamical role in the structure and evolution of the interstellar medium.
However, the issue has not generated too many interests until fairly recently.
For instances, in the past couple decades, efforts have been made on the understanding of the influencing of cosmic rays on instabilities
related to interstellar medium \citep[such as, Parker instability, Jeans instability, magnetorotional instability, see, e.g.][]
{Parker_1966,Kuznetsov_1983,Hanasz_2000,Hanasz_2003,Ryu_2003,Kuwabara_2004,Kuwabara_2006,Ko_2009,Lo_2011,Kuwabara_2015,Heintz_2018,Heintz_2020};
and on cosmic ray modified structures and outflows \citep[e.g., shocks, galactic halo, stellar and galactic winds, see e.g.,][]
{Drury_1981,McKenzie_1982,Ko_1997,Ko_2001,Ghosh_1983,Koetal_1991,Ko_1991a,Ko_1988,Ko_1991b,Ipavich_1975,Zirakashvili_1996,
Everett_2008,Yang_2012,Girichidis_2016,Recchia_2016,Ruszkowski_2017,Mao_2018,Farber_2018,Holguin_2019,Dorfi_2012,Dorfi_2019,Yu_2020}.
Especially, in the past ten years or so, there are a lot of attentions on cosmic ray modified galactic wind.

Cosmic rays driven winds was first discussed by \citet{Ipavich_1975}.
He mentioned the importance of cosmic rays coupling with thermal plasma and hydromagnetic waves.
\citet{Breitschwerdt_1991,Breitschwerdt_1993} studied the role played by cosmic rays on the structure of galactic winds in detail
and concluded that wind can be launched at some distance from the galactic disk.
\citet{Zirakashvili_1996} examined the coupling between cosmic rays and plasma in more detail. They investigated the effect of
Alfv\'en wave damping on heating of thermal gas.
They also studied how the rotation of galactic disk affected the flow.
These works were further extended by \citet{Everett_2008}. They found that both cosmic rays and thermal gas pressure are necessary
to describe the observed soft galactic X-ray and radio emissions.
\citet{Recchia_2016} explained the implications and evolution of the cosmic rays driven wind by solving the hydro dynamical equations
for the wind and transport equation of the cosmic rays under the assumption of the self-generated diffusion, advection with self-excited Alfv\'en waves.
\citet{Wiener_2017,Ruszkowski_2017,Farber_2018} underlined the importance of cosmic ray streaming and its consequences on galactic winds.
\citet{Wiener_2017} discussed in detail the differences between outflows driven by cosmic ray diffusion and advection.
These issues will be addressed in this work but with less elaborate models.
For a summary on the development of galactic wind, the reader is referred to the nice review by \citet[][and references therein]{Zhang_2018}.

In this work we would like to explore possible fluid outflows from a gravitational potential well in the presence of cosmic rays (e.g., galactic wind).
We systematically investigate all possible solutions with specific boundary conditions at the base of the potential well (e.g., properties of ISM),
and examine the influence of cosmic rays.
Cosmic rays interact with magnetized plasma through embedded magnetic irregularities.
In the process they generate waves and exert feedback on the plasma.
The dynamics of the system is best studied under a hydrodynamic model \citep[e.g.,][]{Drury_1981,McKenzie_1982,Ko_1992}.
The most comprehensive hydrodynamic model for cosmic-ray-plasma system is the four-fluid model introduced by \citet{Ko_1992} \citep[see also][]{Zweibel_2017},
which includes thermal plasma, cosmic rays and both forward and backward propagating self-excited Alfv\'en waves.
The cosmic rays and waves are considered as massless fluid but with comparable energy density as the thermal plasma.
The system has several energy exchange mechanisms among the four components.
Even for one-dimensional flow without external potential there exists quite a variety of flow profiles \citep{Ko_2001}.
In this article, we focus mainly on a reduced system: three-fluid or one-wave system (in particular, forward propagating wave),
but still some examples of the four-fluid system will be touched upon.
Basically, our outflow model is similar to the three-fluid model introduced by \citet{Breitschwerdt_1991,Breitschwerdt_1993}.
Yet we do have some additional features that leads to some rather intriguing flow profiles.
For instances, a wave dependent diffusion coefficient of cosmic rays provides the possibility of cosmic rays decoupling from the flow,
and the flow may transit to a pure thermal outflow.
Under the right conditions, wave damping can significantly affect the flow profile.
Several wave damping mechanisms will be studied, including the very special (and kind of artificial) ``completely local wave dissipation''
proposed by \citet{Breitschwerdt_1991}, and other more typical ones, such as, nonlinear Landau damping \citep[e.g.,][]{Kulsrud_2005}.
The paper is organized as follows.
In Section~\ref{sec:model} we present the four-fluid model of the cosmic-ray-plasma system in flux-tube formation.
The four fluids refer to thermal plasma, cosmic ray, forward and backward propagating Alfv\'en waves.
In Section~\ref{sec:without_diffusion}, the one-wave (or three-fluid) without cosmic ray diffusion is studied analytically.
Section~\ref{sec:comparison} shows examples and comparisons of different possible cases, including one-wave systems with and without diffusion
with and without damping, and two-wave systems.
A summary and some concluding remarks are provided in Section~\ref{sec:Summary}.

\section{Hydrodynamic model for cosmic-ray-plasma system}
\label{sec:model}

When cosmic rays propagate through a plasma, they can excite hydromagnetic waves through streaming instabilities.
There are energy exchanges among the thermal plasma, cosmic rays and waves.
If the energy density of the cosmic rays (and waves) are comparable to the thermal plasma (e.g., in the ISM of our Galaxy),
the dynamics of the plasma will be affected by the cosmic rays (and waves).
It is convenient to study the dynamics of the system under a hydrodynamical model.
We adopt a four-fluid model which comprises thermal plasma, cosmic rays, two Alfv\'en waves (forward and backward propagating),
where the cosmic rays and waves are considered as massless fluid but with pressure comparable to the thermal plasma
\citep[][see also \citet{Zweibel_2017}]{Ko_1992}.

It is a formidable task to find general solutions to the full system.
Here we restrict ourselves to a special situation by assuming the magnetic field of the system a prescribed smooth configuration,
and we only consider the dynamics along the magnetic field only. We call this flux-tube formulation.
In this formation, the plasma flow is along the magnetic field and there is no cross-field line diffusion of cosmic ray.
In steady state, the governing equations are:
\begin{equation}\label{eq:bfield}
  B \Delta = \psi_B = \sqrt{\mu_0} \psi_B^\prime \Delta\,,
\end{equation}
\begin{equation}\label{eq:mass}
  \rho U \Delta = \psi_m = \psi_m^\prime \Delta\,,
\end{equation}
\begin{equation}\label{eq:momentum}
  \rho U\frac{d U}{d\xi} = -\frac{d}{d\xi}\left(P_g+P_c+P^+_w+P^-_w\right)-\rho\frac{d\Psi}{d\xi}\,,
\end{equation}
\begin{equation}\label{eq:gas}
  \frac{1}{\Delta}\frac{d F_g \Delta}{d\xi} = U\frac{d P_g}{d\xi} +L^+_w+L^-_w\,,
\end{equation}
\begin{equation}\label{eq:cr}
  \frac{1}{\Delta}\frac{d F_c \Delta}{d\xi} = \left[U+(e_+-e_-)V_A\right]\frac{d P_c}{d\xi}+\frac{P_c}{\tau}\,,
\end{equation}
\begin{equation}\label{eq:waves}
  \frac{1}{\Delta}\frac{d F^\pm_w \Delta}{d\xi} = U\frac{d P^\pm_w}{d\xi}\mp e_\pm V_A\frac{d P_c}{d\xi} - \frac{P_c}{2\tau} - L^\pm_w\,,
\end{equation}
where the energy fluxes are
\begin{equation}
  F_g = \left(E_g+P_g\right)U = \frac{\gamma_g P_g}{\left(\gamma_g-1\right)}\,U\,,
\end{equation}
\begin{align}
  F_c & = \left(E_c+P_c\right)\left[U+(e_+-e_-)V_A\right]-\kappa\frac{d E_c}{d\xi} \nonumber \\
      & = \frac{\gamma_c P_c}{\left(\gamma_c-1\right)}\left[U+(e_+-e_-)V_A\right]-\frac{\kappa}{\left(\gamma_c-1\right)}\frac{d P_c}{d\xi} \,,
\end{align}
\begin{equation}
  F^\pm_w = E^\pm_w\left(U\pm V_A\right)+P^\pm_w U = P^\pm_w\left(3U\pm 2V_A\right)\,.
\end{equation}
Here $\xi$ is the coordinate along the magnetic field,
$\Delta(\xi)$ is the cross-sectional area of the flux tube,
$\rho$ and $U$ are the density and flow velocity of the plasma,
$P_g$, $P_c$ and $P^\pm_w$ are the pressures of thermal plasma, cosmic ray and waves ($\pm$ denote forward and backward propagating Alfv\'en waves).
$\Psi$ is the external gravitational potential.
$\kappa$ is the diffusion coefficient of cosmic rays along magnetic field line,
$P_c/\tau$ and $L^\pm_w$ represent stochastic acceleration and wave damping, respectively.
Equation~(\ref{eq:bfield}) and Equation~(\ref{eq:mass}) come from divergent-free of magnetic field and mass continuity,
and $\psi_B$ and $\psi_m$ are called the magnetic flux and the mass flow rate, respectively.
The terms $\mp e_\pm V_A dP_c/d\xi$ in Equation~(\ref{eq:waves}) are wave excitation by cosmic ray streaming instability.
Thus negative cosmic ray pressure gradient promotes/restrains forward/backward propagating wave and vice versa.
Note that $e_++e_-=1$ and $e_\pm$ can be considered as the weighting of forward and backward waves (they are related to $\nu_\pm$,
the collision frequencies of cosmic rays by forward and backward waves).
The Alfv\'en speed is given by
\begin{equation}\label{eq:VA}
  V_A = \frac{B}{\sqrt{\mu_0\rho}}
  =\frac{\psi_B^\prime\sqrt{\rho}\, U}{\psi_m^\prime}={\tilde\psi}\sqrt{\rho}\, U=M_A^{-1}U\,,
\end{equation}
where $M_A$ is the Alfv\'en Mach number. Here ${\tilde\psi}=\psi_B^\prime/\psi_m^\prime$,
$\psi_m^\prime=\psi_m/\Delta=\rho U$ and $\psi_B^\prime=\psi_B/\sqrt{\mu_0}\Delta=B/\sqrt{\mu_0}$
(Equations~(\ref{eq:bfield}) \& (\ref{eq:mass})).

The system has an useful integral and we call it energy constant (the ratio of total energy flux to mass flux, cf. Bernoulli's equation)
\begin{equation}\label{eq:Bernoulli}
  {\cal E}_{\rm tot} = \frac{U^2}{2}+\frac{\Delta}{\psi_m}\left(F_g+F_c+F^+_w+F^-_w\right)+\Psi \,.
\end{equation}
Moreover, if $L^\pm_w=0$ (and $e_++e_-=1$), we have two more integrals, the entropy integral of the thermal gas
\begin{equation}\label{eq:entropy}
  A_g=P_g\rho^{-\gamma_g}\,,
\end{equation}
and the wave action integral
\begin{equation}\label{eq:wave_action}
  {\cal W}_{\rm A} = \Delta\left[F_c+E^+_w\frac{\left(U+V_A\right)^2}{V_A}-E^-_w\frac{\left(U-V_A\right)^2}{V_A}\right] \,.
\end{equation}

\subsection{Outflow equation for two-wave system}
\label{sec:two_wave_diffusion}

To discuss possible outflow from a gravitational potential well, it is useful to express the momentum equation Equation~(\ref{eq:momentum})
in the form of a wind equation \citep[cf. classic stellar wind, e.g.,][]{Parker_1958}.
In the presence of cosmic ray diffusion, the outflow or wind equation must include the $d P_c/d\xi$ term explicitly,
\begin{align}\label{eq:two_wave_wind}
  & \left(1- M^{-2}_{\rm eff}\right)U\frac{dU}{d\xi} = \frac{a^2_{\rm eff}}{\Delta}\frac{d\Delta}{d\xi}-\frac{d\Psi}{d\xi} \nonumber \\
  &\quad -\frac{1}{\rho}\frac{d P_c}{d\xi}
  \left[\frac{e_+(1+\frac{1}{2}M_A^{-1})}{(1+M_A^{-1})}+\frac{e_-(1-\frac{1}{2}M_A^{-1})}{(1-M_A^{-1})}\right] \nonumber \\
  &\quad +\frac{1}{\rho U}\left[\frac{P_c}{2(1-M_A^{-2})\tau}-(\gamma_g-1)\left(L^+_w+L^-_w\right) \right. \nonumber \\
  &\quad\quad\quad \left. +\frac{L^+_w}{2(1+M_A^{-1})}+\frac{L^-_w}{2(1-M_A^{-1})}\right]\,.
\end{align}
The outflow equation is supplemented by
\begin{align}\label{eq:two_wave_cr}
  & \frac{d}{d\xi}\left(\kappa\frac{d P_c}{d\xi}\right) = -\kappa\frac{d P_c}{d\xi}\left(\frac{1}{\Delta}\frac{d\Delta}{d\xi}\right)
  -\left(\gamma_c-1\right)\frac{P_c}{\tau} \nonumber \\
  &\quad +\frac{U\left[1+(e_+-e_-)M_A^{-1}\right]\rho^{\gamma_c}}{\left|1+(e_+-e_-)M_A^{-1}\right|^{\gamma_c}} \nonumber \\
  &\quad\quad\quad * \frac{d}{d\xi}\left\{P_c\left|\frac{1+(e_+-e_-)M_A^{-1}}{\rho}\right|^{\gamma_c}\right\}\,,
\end{align}
\begin{equation}
  \frac{d P_g}{d\xi} = \frac{\gamma_g P_g}{\rho}\frac{d\rho}{d\xi}+\frac{(\gamma_g-1)}{U}\left(L^+_w+L^-_w\right)\,,
\end{equation}
\begin{align}\label{eq:two_wave_wave}
  & \frac{d P^\pm_w}{d\xi} = \frac{3 P^\pm_w}{2\rho}\frac{(1\pm\frac{1}{3}M_A^{-1})}{(1\pm M_A^{-1})}\frac{d\rho}{d\xi}
  \mp\frac{e_\pm M_A^{-1}}{2(1\pm M_A^{-1})}\frac{d P_c}{d\xi} \nonumber \\
  &\quad -\frac{P_c}{4U(1\pm M_A^{-1})\tau}-\frac{L^\pm_w}{2U(1\pm M_A^{-1})}\,.
\end{align}
Here the effective sound speed is
\begin{equation}\label{eq:sound_speed_2}
  a^2_{\rm eff} = a^2_g+a^{+\,2}_w\frac{\left(1+\frac{1}{3}M_A^{-1}\right)}{\left(1+M_A^{-1}\right)}
  +a^{-\,2}_w\frac{\left(1-\frac{1}{3}M_A^{-1}\right)}{\left(1-M_A^{-1}\right)}\,.
\end{equation}
where
\begin{equation}\label{eq:sound_speeds}
  a^2_g = \frac{\gamma_g P_g}{\rho}\,,
  \quad a^{+\,2}_w = \frac{3 P^+_w}{2\rho}\,,
  \quad a^{-\,2}_w = \frac{3 P^-_w}{2\rho}\,.
\end{equation}
As long as diffusion is included then the solutions of the system (whether it is a two-wave, or one-wave, or even ``no-wave'' system)
cannot be presented simply in the plane of $(\xi,U)$ as in the classical stellar wind case.
${\cal E}_{\rm tot}$ is not a function of $\xi$ and $U$ only, and the sonic point or critical point becomes a line or plane
depending on the running variables \citep{Ko_1987}.

The problem is simpler if there is only one wave in the system as the stochastic acceleration term vanishes automatically.
Moreover, $\{e_+,e_-\}=\{1,0\}$ for forward propagating wave and $\{e_+,e_-\}=\{0,1\}$ for backward propagating wave.
It is particular simple if cosmic ray diffusion can be neglected and the wave damping term $L^\pm_{w}$ is of some specific form (see below).
The problem can then be reduced to a problem similar to the classic stellar wind problem \citep{Parker_1958}.
In Section~\ref{sec:without_diffusion}, we consider two specific forms of wave damping in one-wave systems without cosmic ray diffusion.
The first one is no wave damping $L^\pm_w=0$.
The second one is the damping is just strong enough to negate the excitation by streaming instability $L^\pm_w=\mp e_\pm V_A\, dP_c/d\xi$
\citep[e.g.,][]{Breitschwerdt_1991,Everett_2008,Heintz_2018}.
This is called completely local wave dissipation (CLWD) \citep[][]{Breitschwerdt_1991}.
We point out that $L^\pm_w$ may not be positive definite in this model, i.e., it may act as a sink or a source.

\section{One wave system without cosmic ray diffusion}
\label{sec:without_diffusion}

Without diffusion in one-wave system the cosmic ray pressure can be written in terms of density analytically.
Putting either $\{e_+,e_-\}=\{1,0\}$ or $\{0,1\}$, and $\kappa=0$ ($1/\tau=0$ for one-wave system) into Equation~(\ref{eq:cr}), we get
\begin{equation}\label{eq:pc_no_diff}
  P_c\left|\frac{\rho}{1\pm M_A^{-1}}\right|^{-\gamma_c} = A_c\,,
\end{equation}
where the inverse Alfv\'en Mach number $M_A^{-1}={\tilde\psi}\sqrt{\rho}$ (see, Equation~(\ref{eq:VA})).
Here and the following, the upper/lower sign corresponds to forward/backward propagating wave system.

\subsection{No wave damping case}
\label{sec:no_diff_no_damp}
If there is no wave damping ($L^\pm_w=0$), then Equation~(\ref{eq:gas}) gives
\begin{equation}
  P_g\rho^{-\gamma_g} = A_g\,.
\end{equation}
With the expression of $P_c$ in Equation~(\ref{eq:pc_no_diff}), Equation~(\ref{eq:waves}) gives
(this is basically the wave action integral),
\begin{equation}
  \left[P^\pm_w\pm\frac{\gamma_c P_c}{2(\gamma_c-1)}\frac{M_A^{-1}}{\left(1\pm M_A^{-1}\right)}\right]
  \frac{\left(1\pm M_A^{-1}\right)^2}{\rho^{3/2}} = A^\pm_w\,.
\end{equation}
Thus all the pressures can be expressed as a function of density.

\subsection{Completely local wave dissipation case}
\label{sec:no_diff_with_CLWD}

For the case of complete local wave dissipation,
$L^\pm_w=\mp V_A dP_c/d\xi$, and Equation~(\ref{eq:waves}) gives
\begin{equation}
  P^\pm_w\frac{\left(1\pm M_A^{-1}\right)^2}{\rho^{3/2}} = A^\pm_w\,.
\end{equation}
and the thermal pressure can be expressed as
\begin{equation}
  P_g\rho^{-\gamma_g}-{\cal F}^\pm(\rho) = A_g\,,
\end{equation}
where
\begin{align}
  & {\cal F}^\pm(\rho) = \mp\frac{A_c\gamma_c(\gamma_g-1)M_A^{-1}}{(2\gamma_c-2\gamma_g+1)}\rho^{\gamma_c-\gamma_g} \nonumber \\
  & \ \ *\left[ _2F_1\left(\gamma_c,2\gamma_c-2\gamma_g+1;2\gamma_c-2\gamma_g+2;\mp M_A^{-1}\right) \right. \nonumber \\
  & \quad \left. +_2F_1\left(\gamma_c+1,2\gamma_c-2\gamma_g+1;2\gamma_c-2\gamma_g+2;\mp M_A^{-1}\right) \right]\,,
\end{align}
and $_2F_1(a,b;c;z)$ is the hypergeometric function. For example, if $\gamma_g=5/3$, $\gamma_c=4/3$, then
\begin{align}
  {\cal F}^\pm(\rho) & = \mp\frac{2}{3}A_c M_A^{-1}\rho^{-1/3}\left|1\pm M_A^{-1}\right|^{-4/3}\left(8\pm 7M_A^{-1}\right) \nonumber \\
                     & = \mp\frac{2}{3}P_c M_A^{-1}\rho^{-5/3}\left(8\pm 7M_A^{-1}\right)\,.
\end{align}

\subsection{Outflow equation for one-wave system without cosmic ray diffusion}
\label{sec:wind_eq}

Since all the pressures ($P_g$, $P_c$, $P^\pm_w$) can be expressed as functions of density $\rho$ which in turn can be
expressed as a function of $U$ and $\xi$ (Equation~(\ref{eq:mass})), the problem is similar to the classic stellar wind problem
\citep[][see also \citep{Bondi_1952}]{Parker_1958,Parker_1963}. The solution is given by the Bernoulli's equation Equation~(\ref{eq:Bernoulli})
(with only one wave energy flux)
which gives a relation between $U$ and $\xi$ for a particular set of parameters
$\{\psi_B,\psi_m,A_g,A_c,A^\pm_w,{\cal E}_{\rm tot}\}$.
To understand the qualitative behaviour of the solutions, we rely on the outflow or wind equation.
For one-wave system without cosmic ray diffusion, the outflow equation is (cf. Equation~(\ref{eq:two_wave_wind})),
\begin{align}\label{eq:wind_no_diff}
  & \left(1- M^{-2}_{\rm eff}\right)U\frac{dU}{d\xi} = \frac{a^2_{\rm eff}}{\Delta}\frac{d\Delta}{d\xi}-\frac{d\Psi}{d\xi} \nonumber \\
  &\quad\quad\quad +\frac{L^\pm_w}{\rho U}\left[\frac{1}{2(1\pm M_A^{-1})}-(\gamma_g-1)\right]\,,
\end{align}
where the effective Mach number $M_{\rm eff}=U/a_{\rm eff}$.
Here $L^\pm_w$ represents wave damping mechanisms other than CLWD.
The effective sound speed $a_{\rm eff}$ is given by
\begin{align}\label{eq:total_sound}
  a^2_{\rm eff} & = a^2_g+a^{\pm\,2}_w\frac{\left(1\pm\frac{1}{3}M_A^{-1}\right)}{\left(1\pm M_A^{-1}\right)} \nonumber \\
                & \quad +a^2_c\left\{q\left[1\mp(\gamma_g-1)M_A^{-1}\right]
                \frac{\left(1\pm\frac{1}{2}M_A^{-1}\right)}{\left(1\pm M_A^{-1}\right)} \right. \nonumber \\
                & \quad\quad\quad\quad \left. +(1-q)\frac{\left(1\pm\frac{1}{2}M_A^{-1}\right)^2}{\left(1\pm M_A^{-1}\right)^2}\right\}\,,
\end{align}
and
\begin{equation}\label{eq:sound_speeds2}
  a^2_g = \frac{\gamma_g P_g}{\rho}\,,
  \quad a^2_c = \frac{\gamma_c P_c}{\rho}\,,
  \quad a^{\pm\,2}_w = \frac{3 P^\pm_w}{2\rho}\,,
\end{equation}
where $q=1$ for CLWD and $q=0$ for no wave damping or other damping mechanisms (NLLD, INF).
We are interested in outflows against gravity, i.e., $U>0$ and $-d\Psi/d\xi\le 0$.
Even if $-d\Psi/d\xi\le 0$, the RHS of Equation~(\ref{eq:wind_no_diff}) can be positive for divergent flux tube (i.e., $d\Delta/d\xi>0$).
When solution passes through the curve defined by the RHS of Equation~(\ref{eq:wind_no_diff}) equals zero, $dU/d\xi=0$;
and when solution passes through the curve defined by $M_{\rm eff}=1$, $dU/d\xi\rightarrow\infty$ (i.e., unphysical).
The intersection point(s) of the two curves, if exists, is called the fixed point, critical point, or in our case the sonic point.

\begin{figure*}[htb]
\centering
\vskip -7cm
\includegraphics[width=0.85\textwidth]{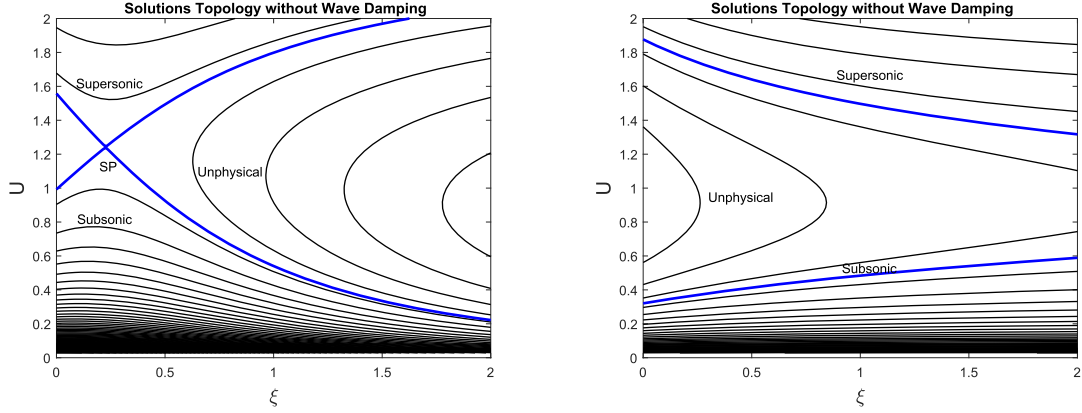}
\vskip -7cm
\caption{
Solution curves expressed as velocity against position for the case of one-wave system without
cosmic rays diffusion and without wave damping.
In this figure, $\gamma_g=5/3$, $\gamma_c=4/3$, $GM=1$, $a=1$,
$\psi_B=1$, $\psi_m=1.2$,
$A_g=0.5$, $A_c=0.5$, $A^+_w=0.5$.
{\it Left}: In a divergent flux tube geometry with $d=2$ and $\xi_d=1$.
{\it Right}: In a constant flux tube geometry (or one-dimensional geometry) with $d=0$ and $\xi_d=1$.
The solutions are divided into subsonic branch, supersonic branch and unphysical regions
(regions where two different velocities occur at the same location in one solution).
The blue curves are called transonic solutions.
In the divergent flux tube case (left panel) the two transonic solutions pass through the sonic point,
while in the constant flux tube case (right panel) the two transonic solutions will meet at the sonic point at infinity.
}
\label{fig:contourplot}
\end{figure*}

For illustration, let us take
\begin{equation}\label{eq:tube_potential}
  \Delta(\xi) = \Delta_0\left[1+\left(\frac{\xi}{\xi_d}\right)^d\right]\,,
  \quad \Psi = -\frac{GM}{\left(a+\xi\right)}\,.
\end{equation}
We plot the solution curves of forward propagating wave systems without wave damping with
$d=2$ (a divergent flux tube geometry)
in the left panel of Figure~\ref{fig:contourplot}
and
$d=0$ (i.e., constant flux tube of one-dimensional geometry)
in the right panel of Figure~\ref{fig:contourplot}.
In the figures, we set $\gamma_g=5/3$, $\gamma_c=4/3$, $a=1$, $\xi_d=1$.
Furthermore,
$\{\psi_B,\psi_m,A_g,A_c,A^+_w\}=\{1, 1.2, 0.5, 0.5, 0.5\}$
in both panels.
Different curves correspond to different energy constant ${\cal E}_{\rm tot}$.
We also note that for each ${\cal E}_{\rm tot}$ there are two solution curves, one subsonic and one supersonic.
In the left panel of Figure~\ref{fig:contourplot},
the two curves passes through the sonic point are called transonic solutions
(blue curves in the figure).
${\cal E}_{\rm tot}$ increases from the unphysical regions to the subsonic and supersonic regions,
and the surface of ${\cal E}_{\rm tot}$ (as a function of $(\xi,U)$) resembles a saddle (and in fact, the sonic point is a saddle point).
The same applies to
the right panel of Figure~\ref{fig:contourplot}
except that the sonic point is at infinity.

\begin{figure*}[htb]
\centering
\includegraphics[width=0.8\textwidth]{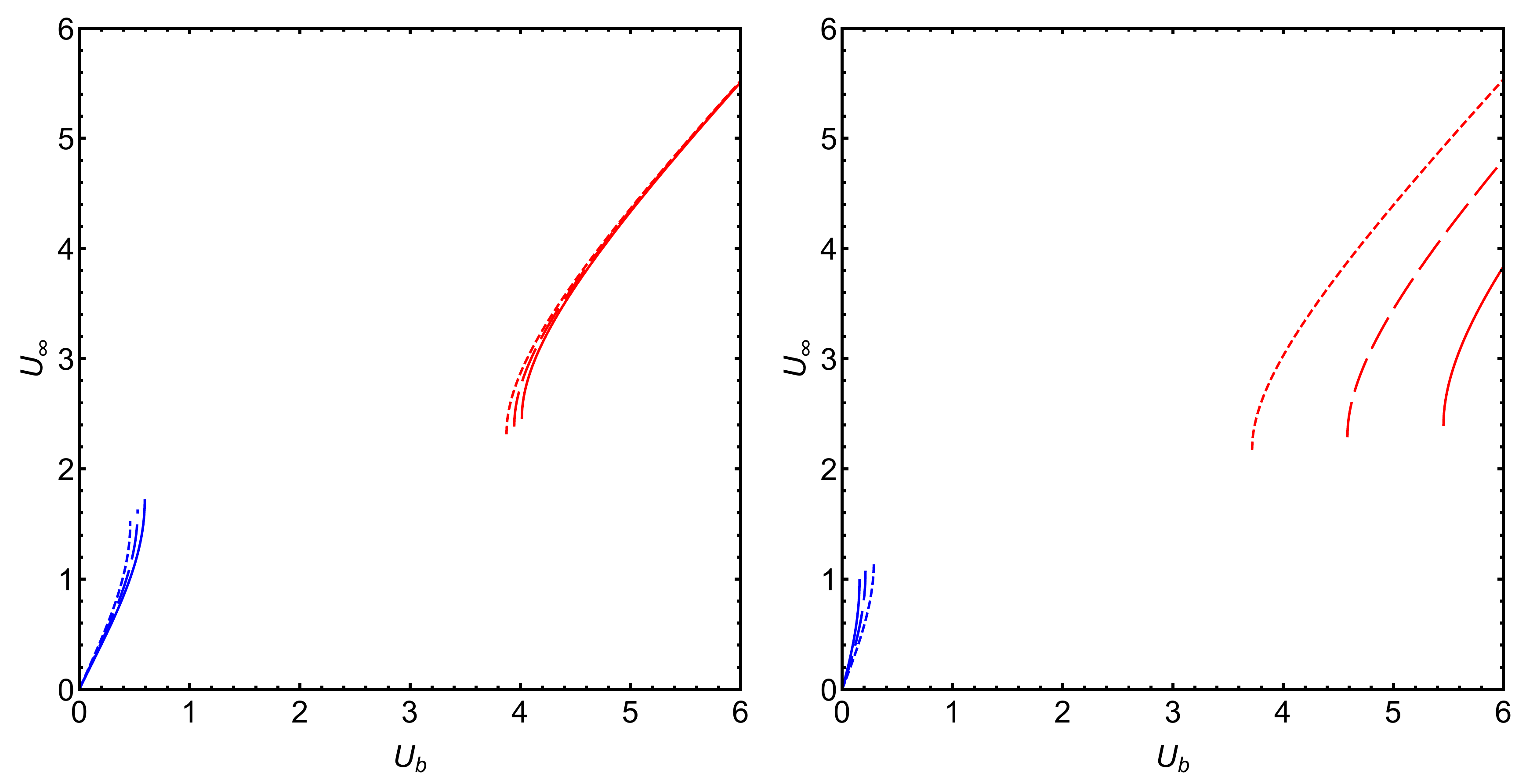}
\caption{
Asymptotic velocity versus velocity at the base of the potential well. The subsonic branch is blue and the supersonic branch is red.
The parameters used in both panels are
$\gamma_g=5/3$, $\gamma_c=4/3$, $d=0$, $\xi_d=1$, $\Psi_b=-2.4$ and $\psi_B^\prime=B_b/\sqrt{\mu_0}=1$.
{\it Left}: The case of one-wave system (i.e., three-fluid model) without diffusion and without wave damping
(see Sections~\ref{sec:wind_eq} \& \ref{sec:compare_1}).
The boundary conditions are $\rho_b=1$, $P_{gb}=2$, $P_{cb}=0.7$.
{\it Right}: The case of quasi-thermal outflow without wave damping (see Section~\ref{sec:quasi-thermal}).
The boundary conditions are $\rho_b=1$, $P_{gb}=2$, $P^-_{wb}=0$.
In both panels, the dashed, long-dashed and solid lines correspond to forward wave pressure at the base
$P^+_{wb}=0,\,0.25,\,0.5$,
respectively.
}
\label{fig:asymptotic}
\end{figure*}

In passing we mention that the corresponding figures for CLWD damping look very similar to
Figure~\ref{fig:contourplot}.
As shown in the right panel of Figure~\ref{fig:contourplot},
the velocity profile in the subsonic (supersonic) branch increases (decreases) monotonically
from the boundary value at the base of the potential well to some asymptotic value at large distances.
Clearly, the asymptotic velocity depends on the boundary value of various quantities at the base of the potential well, such as,
$U_b$, $\rho_b$, $P_{gb}$, $P_{cb}$, $P^+_{wb}$, $B_b$, $\Psi_b$, etc.
To illustrate the idea, suppose an outflow is launched at a height 1 kpc above the mid-plane of our Galaxy at the position of the Sun.
From \citet{Ferrire_2001}, at this height the density and pressure of hot gas are $\sim 0.0065$ cm$^{-3}$ and $\sim 2\times 10^{-12}$ erg cm$^{-3}$,
the pressures of cosmic rays and magnetic field are $\sim 0.7\times 10^{-12}$ erg cm$^{-3}$ and $\sim 0.5\times 10^{-12}$ erg cm$^{-3}$, respectively.
We adopt the gravitational potential model of the disk by \citet{Barros_2016} (see Table~3 of their paper).
We take the galactodistance of the Sun as $\sim 8.5$ kpc.
If we pick the units for pressure, density and velocity as $10^{-12}$ erg cm$^{-3}$, $10^{-26}$ g cm$^{-3}$ and $100$ km s$^{-1}$, respectively,
then $\rho_b=1$, $P_{gb}=2$, $P_{cb}=0.7$, $\psi_B^\prime=B_b/\sqrt{\mu_0}=1$, $\Psi_b=-2.4$.
We plot the asymptotic velocity $U_\infty$ against the velocity at the base $U_b$
in the left panel of Figure~\ref{fig:asymptotic}.
In the figure, the dashed, long-dashed and solid lines correspond to
$P^+_{wb}=0,\,0.25,\,$0.5, respectively.
The blue ones belong to subsonic branch and the red ones supersonic branch.
The region between them corresponds to those values of velocity at the base that give unphysical solutions.
This region will be larger if the potential well is deeper.

We point out that the case of backward propagating wave is more complicated as solutions for super-Alfv\'enic flow
and sub-Alfv\'enic flow can be qualitatively different (as indicated by the lower sign in the above expressions).
For instance, it can be shown from Equation~(\ref{eq:waves}) that the contribution of the wave damping term
$L^-_w$ to $d P^-_w/d\xi$ is positive for sub-Alfv\'enic flow.
Note also that the energy flux of backward propagating wave is in the negative $\xi$-direction if $M_A<2/3$.

In the following discussions on one-wave systems, we focus on forward propagating wave only.
Moreover, we consider super-Alfv\'enic flow only.


\begin{deluxetable*}{ccccccccc}
\tablewidth{0pt}
\tablecaption{\label{tab:P_table}
Parameters in Figures~\ref{fig:3f_nodiffusion}--\ref{fig:quasi_thermal_WDNLLD}.
$\gamma_g=5/3$, $\gamma_c=4/3$, $GM=1$, $a=1$,
${\psi_B^\prime}=1$.
}
\tablehead{
\colhead{Figure} &\colhead{Branch} & \colhead{$U_b$} & \colhead{$P_{gb}$} & \colhead{$P_{cb}$}
& \colhead{$P^+_{wb}$} & \colhead{$P^-_{wb}$} & \colhead{NLLD ($\delta^\pm$)} & \colhead{Diffusion ($\kappa$)} \\
}
\startdata
\ref{fig:3f_nodiffusion} & Sub & 1.2 & 3 & 1 & 0.25 & 0 & 1 & 0 \\
   & Super & 4 & 3 & 1 & 0.25 & 0 & 1 & 0\\
\hline
\ref{fig:3f_diffusion} & Sub  & 1.2 & 3 & 1 & 0.25 & 0 & 1 & $\kappa^*$ \\
   & Super & 4 & 3 & 1 & 0.25 & 0 & 1 & $\kappa^*$ \\
\hline
\ref{fig:3f_4f_no_damp} & Sub  & 1.2 & 3 & 1 & 0.25 & 0, 0.125 & 0 & 0, $\kappa^*$, $\kappa^{**}$ \\
   & Super  & 4 & 3 & 1 & 0.25 & 0, 0.125 & 0 & 0, $\kappa^*$, $\kappa^{**}$ \\
\hline
\ref{fig:3f_4f_NLLD} & Sub  & 1.2 & 3 & 1 & 0.25 & 0, 0.125 & 1 & 0, $\kappa^*$, $\kappa^{**}$  \\
   & Super & 4 & 3 & 1 & 0.25 & 0, 0.125 & 1 & 0, $\kappa^*$, $\kappa^{**}$ \\
\hline
\ref{fig:quasi_thermal_ND} & Sub  & 1.2 & 5 & 1 & 0.25 & 0 & 0 & 0, $\kappa^*$  \\
   & Super & 4 &1 & 1 & 0.25 & 0 & 0 & 0, $\kappa^*$ \\
\hline
\ref{fig:quasi_thermal_WDNLLD} & Sub  & 1.2 & 5 & 1 &  0.25 & 0 & 0, 1, 2 & $\kappa^*$  \\
   &     & 1.2 & 5 & 0.2, 0.5, 1 &  0.25 & 0 & 1 & $\kappa^*$ \\
   &     & 1.2 & 5 & 1 &  0.15, 0.25, 0.35 & 0 & 1 & $\kappa^*$ \\
   & Super & 4 & 1 & 1 & 0.25 & 0 & 0, 1, 2 & $\kappa^*$  \\
   &     & 4 & 1 & 0.2, 0.5, 1 &  0.25 & 0 & 1 & $\kappa^*$  \\
   &     & 4 & 1 & 1 &  0.15, 0.25, 0.35 & 0 & 1 & $\kappa^*$ \\
\enddata
\tablecomments{
$3\kappa^*=1/P^+_w$ (for one-wave system); $3\kappa^{**} = 1/(P^+_w + P^-_w)$ (for two-wave system).
}
\end{deluxetable*}

\section{Possible cases}
\label{sec:comparison}

Outflows with cosmic rays and waves can be considered in models of different sophistication,
from one-wave system without diffusion and without wave damping to two-wave system with diffusion and wave damping.
In this section, we compare results of different models with the same boundary conditions at the base of the potential well,
e.g., the conditions at the galactic disk for outflow into the galactic halo.
Although there are many cases worth discussing, for simplicity, we restrict ourselves mainly to constant flux tube geometry.
Physically allowable solutions refer to those have finite velocity and pressures at large distances.
We compare one-wave system (specifically, forward propagating wave system) without cosmic ray diffusion and with diffusion,
without wave damping and with wave damping, for both subsonic branch and supersonic branch.
Moreover, we also compare results of two-wave system to one-wave system.

For two-wave systems, we adopt the following working model for $e_\pm$, $\kappa$ and $1/\tau$
\citep[e.g.,][]{Skilling,Ko_1992},
\begin{align}\label{eq:epmkappatau}
  e_\pm & = \frac{\nu_\pm}{(\nu_++\nu_-)}\,, \nonumber \\
  \kappa & = \frac{c^2}{3(\nu_++\nu_-)}\,, \\
  \frac{1}{\tau} & = \frac{16 \nu_+\nu_- V_A^2}{(\nu_++\nu_-)c^2}\,, \nonumber
\end{align}
where $\nu_\pm$ are the collision frequencies of cosmic rays by forward and backward propagating waves.
We take $\nu_\pm\propto P^\pm_w$.

Besides CLWD damping described in Section~\ref{sec:no_diff_with_CLWD},
one can consider other wave damping mechanisms, such as ion-neutral friction (INF) or nonlinear Landau damping (NLLD).
Approximately, in terms of the present variables, $L^\pm_w \propto P_w^\pm\sqrt{\rho P_g}$ for INF,
and $L^\pm_w \propto P_w^{\pm\,2}\sqrt{P_g/\rho B^2}$ for NLLD \citep[e.g.,][]{Kulsrud_2005}.

In the following we will discuss a number of cases in three categories.
For convenience, we summarise the parameter sets corresponding to different cases
(Figures~\ref{fig:3f_nodiffusion}--\ref{fig:quasi_thermal_WDNLLD}) in Table~\ref{tab:P_table}.

\subsection{One-wave systems without diffusion}
\label{sec:compare_1}

First, let us consider typical cases of one forward propagating wave system without cosmic ray diffusion, and in constant flux tube geometry
(i.e., $d=0$ in Equation~(\ref{eq:tube_potential})).
In general physically allowable solutions can be divided into two branches: subsonic and supersonic
(see the right panel of Figure~\ref{fig:contourplot}).
We would like to compare velocity and pressure profiles of different cases (say, with and without wave damping) of the same boundary conditions,
$\{U_b,P_{gb},P_{cb},P^+_{wb},\rho_b,B_b\}$ at $\xi_b$ (the base of the potential well).
Thus they have the same
mass flow rate and magnetic flux,
$\psi_m$ and $\psi_B$
(note that ${\tilde\psi}=\psi_B/\sqrt{\mu_0}\psi_m=\psi_B^\prime/\psi_m^\prime$),
and also the same energy constant ${\cal E}_{\rm tot}$.

\begin{figure*}[htb]
\centering
\vskip -7.5cm
\includegraphics[width=0.9\textwidth]{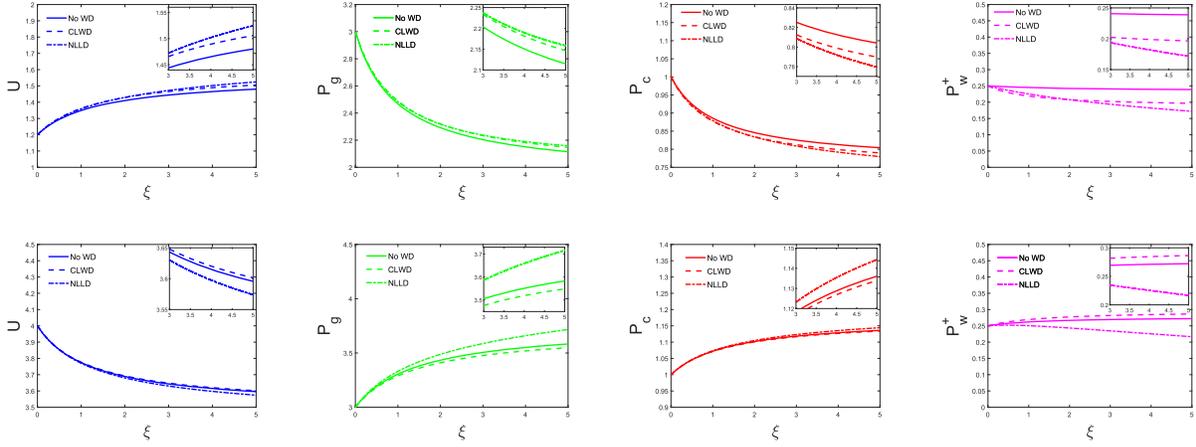}
\vskip -7.5cm
\caption{
Profiles of outflow of one-wave system without cosmic ray diffusion.
In this example, $\gamma_g=5/3$, $\gamma_c=4/3$, $GM=1$, $a=1$ and ${\psi_B^\prime}=1$.
$\delta^+=1$ for nonlinear Landau damping (NLLD).
{\it Upper row}: Subsonic branch with boundary conditions $\psi_m^\prime=1.2$, $U_b=1.2$, $P_{gb}=3$, $P_{cb}=1$, $P^+_{wb}=0.25$.
{\it Lower row}: Supersonic branch with boundary conditions $\psi_m^\prime=4$, $U_b=4$, $P_{gb}=3$, $P_{cb}=1$, $P^+_{wb}=0.25$.
The columns from left to right represent velocity, thermal pressure, cosmic ray pressure and forward propagating wave pressure, respectively.
Solid line, dashed line and dot-dashed line represent no damping (No WD), completely local wave dissipation (CLWD)
and non linear Landau damping (NLLD), respectively.
Some of the profiles closely resemble each other and almost overlap.
The insets are zoom-in versions of the figures in order to make the differences more discernible.
}
\label{fig:3f_nodiffusion}
\end{figure*}

It can be shown from Equation~(\ref{eq:wind_no_diff}) \& (\ref{eq:total_sound}) that when compare with the no wave damping case we have the following:
\begin{itemize}
  \item if $\gamma_g>3/2$ for super-Alfv\'enic regime ($M_A>1$), or if $\gamma_g>5/4$ for sub-Alfv\'enic regime ($M_A<1$):
  \begin{itemize}
    \item subsonic branch: CLWD and other wave damping mechanisms (e.g., INF, NLLD) increase $U$ \& $P_g$ and decrease $P_c$ \& $P^+_w$;
    \item supersonic branch: CLWD increases $U$ \& $P^+_w$ and decreases $P_g$ \& $P_c$,
    while other wave damping mechanisms decrease $U$ \& $P^+_w$ and increase $P_g$ \& $P_c$;
  \end{itemize}
  \item if $\gamma_g<5/4$ for super-Alfv\'enic regime, or if $\gamma_g$ is close enough to one for sub-Alfv\'enic regime:
  \begin{itemize}
    \item subsonic branch: CLWD and other wave damping mechanisms decrease $U$ \& $P^+_w$ and increase $P_g$ \& $P_c$;
    \item supersonic branch: CLWD decreases $U$ \& $P_g$ and increases $P_c$ \& $P^+_w$,
    while other wave damping mechanisms increase $U$ \& $P_g$ and decrease $P_c$ \& $P^+_w$.
  \end{itemize}
\end{itemize}
Figure~\ref{fig:3f_nodiffusion} shows the cases of $\gamma_g=5/3$.
The upper row of the figure is the subsonic branch and the lower row is the supersonic branch.
In each row the profiles of velocity and pressures of three cases are shown: (1) without wave damping $L^+_w=0$ (No WD, solid line),
(2) with completely local wave dissipation $L^+_w=-V_A\, dP_c/d\xi$ (CLWD, dashed line),
and (3) with
nonlinear Landau damping $L^\pm_w \propto P_w^{\pm\,2}\sqrt{P_g/\rho B^2}$ (NLLD, dot-dashed line).
We choose $GM=1$ and $a=1$ in Equation~(\ref{eq:tube_potential})) for the potential, and pick $\xi_b=0$ as the base of the potential well.

As shown in the figure, the effect of CLWD on the profiles of velocity and pressures (in comparison with those of no damping case) is not that much,
in particular for the supersonic branch.
On the other hand,
NLLD has a comparably larger effect
and could significantly alters the profile of $P^+_w$.
Of course, if we pick a smaller $\delta^+$, the effect will be reduced proportionally.
Moreover, if we use INF $L^+_w=\delta^+\, P_w^{+\,}\sqrt{P_g \rho}$ (or even constant damping rate, $L^+_w=\delta^+\, P^+_w$),
the profiles are more or less similar to those ofNLLD.

\begin{figure*}[htb]
\centering
\vskip -7.5cm
\includegraphics[width=0.9\textwidth]{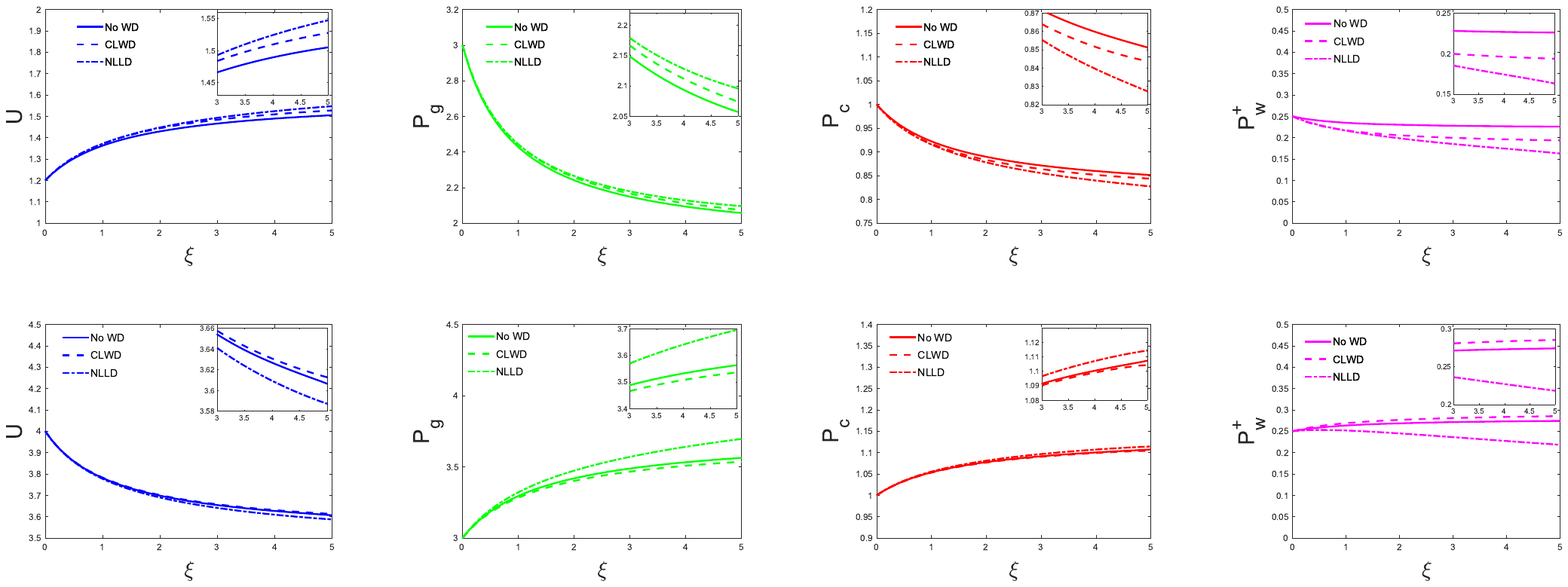}
\vskip -7.5cm
\caption{
Profiles of outflow of one-wave system with cosmic ray diffusion.
In this example, $\kappa=1/(3P^+_w)$, and $\gamma_g=5/3$, $\gamma_c=4/3$, $GM=1$, $a=1$ and ${\psi_B^\prime}=1$.
$\delta^+=1$ for NLLD.
{\it Upper row}: Subsonic branch with boundary conditions $\psi_m^\prime=1.2$, $U_b=1.2$, $P_{gb}=3$, $P_{cb}=1$, $P^+_{wb}=0.25$.
{\it Lower row}: Supersonic branch with boundary conditions $\psi_m^\prime=4$, $U_b=4$, $P_{gb}=3$, $P_{cb}=1$, $P^+_{wb}=0.25$.
The columns from left to right represent velocity, thermal pressure, cosmic ray pressure and forward propagating wave pressure, respectively.
Solid line, dashed line and dot-dashed line represent no damping, completely local wave dissipation
and non linear Landau damping, respectively.
The insets are zoom-in versions of the figures in order to make the differences in profiles more discernible.
}
\label{fig:3f_diffusion}
\end{figure*}

\begin{figure*}[htb]
\centering
\vskip -7.5cm
\includegraphics[width=0.9\textwidth]{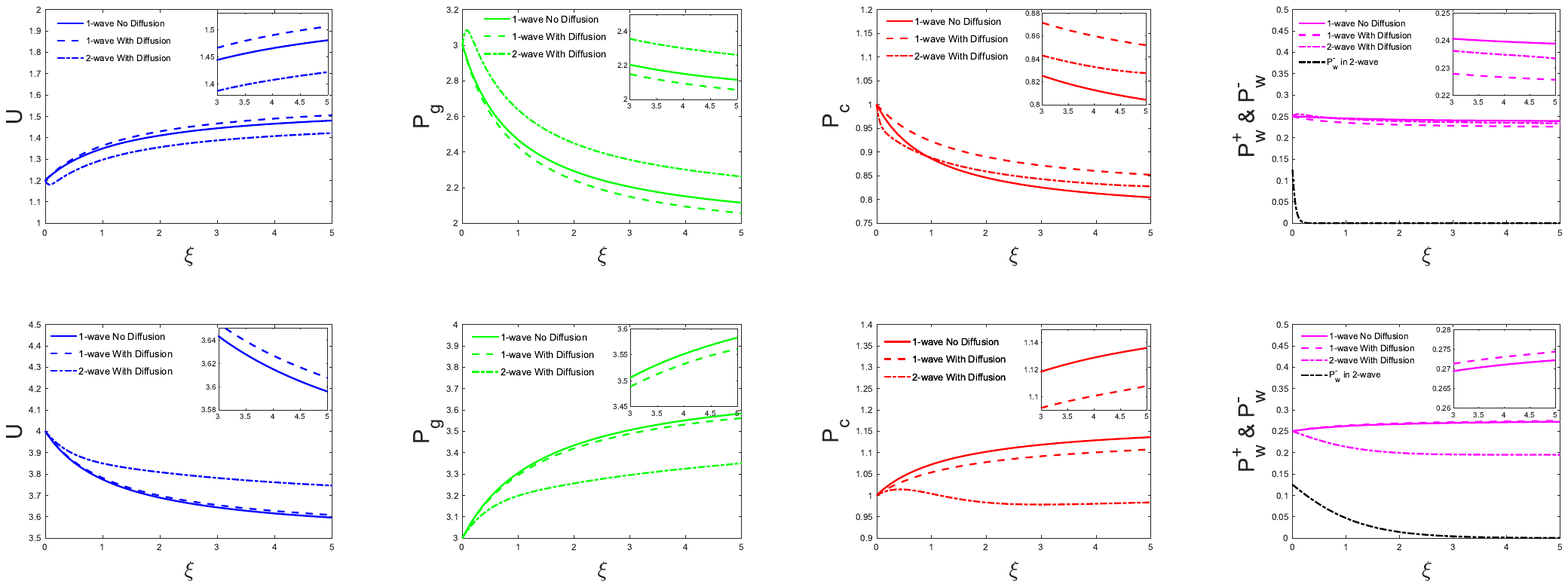}
\vskip -7.5cm
\caption{
This is an example of different systems without wave damping.
Here solid line, dashed line and dot-dashed line represent one-wave system without diffusion, one-wave system with diffusion
and two-wave system with diffusion, respectively.
The black dot-dashed curve in the rightmost column denotes backward propagating wave pressure for the two-wave system.
In this example, $\gamma_g=5/3$, $\gamma_c=4/3$, $GM=1$, $a=1$ and ${\psi_B^\prime}=1$.
{\it Upper row}: Subsonic branch with boundary conditions $\psi_m^\prime=1.2$, $U_b=1.2$, $P_{gb}=3$, $P_{cb}=1$, $P^+_{wb}=0.25$;
$P^-_{wb}=0$ for the one-wave systems and $P^-_{wb}=0.125$ for the two-wave system.
{\it Lower row}: Supersonic branch with boundary conditions $\psi_m^\prime=4$, $U_b=4$, $P_{gb}=3$, $P_{cb}=1$, $P^+_{wb}=0.25$;
$P^-_{wb}=0$ for the one-wave systems and $P^-_{wb}=0.125$ for the two-wave system.
The insets are zoom-in versions of the figures in order to make the differences in profiles more discernible.
}
\label{fig:3f_4f_no_damp}
\end{figure*}

\begin{figure*}[htb]
\centering
\vskip -7.5cm
\includegraphics[width=0.9\textwidth]{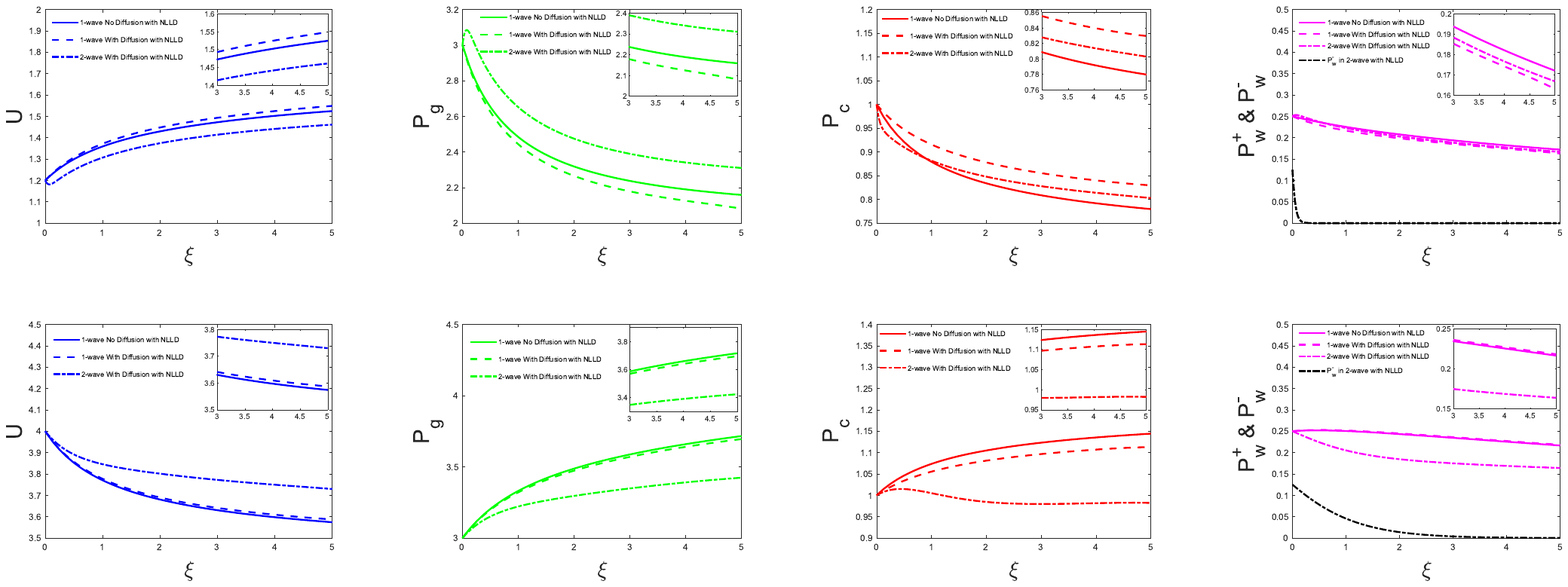}
\vskip -7.5cm
\caption{
This is an example of different systems with nonlinear Landau damping.
Here solid line, dashed line and dot-dashed line represent one-wave system without diffusion, one-wave system with diffusion
and two-wave system, respectively.
The black dot-dashed curve in the rightmost column denotes backward propagating wave pressure for the two-wave system.
In this example, $\gamma_g=5/3$, $\gamma_c=4/3$, $GM=1$, $a=1$, ${\psi_B^\prime}=1$ and $\delta^\pm=1$.
{\it Upper row}: Subsonic branch with boundary conditions $\psi_m^\prime=1.2$, $U_b=1.2$, $P_{gb}=3$, $P_{cb}=1$, $P^+_{wb}=0.25$;
$P^-_{wb}=0$ for the one-wave systems and $P^-_{wb}=0.125$ for the two-wave system.
{\it Lower row}: Supersonic branch with boundary conditions $\psi_m^\prime=4$, $U_b=4$, $P_{gb}=3$, $P_{cb}=1$, $P^+_{wb}=0.25$;
$P^-_{wb}=0$ for the one-wave systems and $P^-_{wb}=0.125$ for the two-wave system.
The insets are zoom-in versions of the figures in order to make the differences in profiles more discernible.
}
\label{fig:3f_4f_NLLD}
\end{figure*}

For the convenience of discerning the slight differences between profiles of different cases (especially the supersonic branch),
we provide enlarged versions of part of the profiles as insets in Figure~\ref{fig:3f_nodiffusion}
(and Figures~\ref{fig:3f_diffusion}--\ref{fig:quasi_thermal_WDNLLD} as well).

\subsection{Cosmic ray accompanied outflows}
\label{sec:compare_2}

When cosmic ray diffusion cannot be neglected, one more condition is needed besides the boundary conditions
$\{U_b,P_{gb},P_{cb},P^+_{wb},\rho_b,B_b\}$ mentioned in Section~\ref{sec:compare_1}.
We are only interested in physically allowable solutions, thus the extra condition is
simply the requirement of finite velocity and pressures at large distances.
This requirement imposes restrictions on the cosmic ray pressure gradient (or the cosmic ray diffusive flux) at $\xi_b$.
It turns out that, depending on the parameters, the requirement has two rather distinct possible outcomes.
We will discuss this two categories in separate subsections.
In the first category (which we call cosmic ray accompanied ouflow will be studied in this subsection),
the requirement demands a specific value of cosmic ray pressure gradient at $\xi_b$
for a given set of $\{U_b,P_{gb},P_{cb},P^+_{wb},\rho_b,B_b\}$. The resulting solution is unique.
In the second category (which we call quasi-thermal outflow will be discussed in Section~\ref{sec:quasi-thermal}),
the requirement allows a range of cosmic ray pressure gradient at $\xi_b$
for a given set of $\{U_b,P_{gb},P_{cb},P^+_{wb},\rho_b,B_b\}$.
As a result, there is a set of solutions satisfying the same given set of boundary conditions $\{U_b,P_{gb},P_{cb},P^+_{wb},\rho_b,B_b\}$.

Once again we would like to compare results of different cases (with and without diffusion, or with and without wave damping).
Now the cosmic ray pressure gradient at $\xi_b$ will not be the same for different cases.
While $\psi_m$ and $\psi_B$ (or ${\tilde\psi}$) are the same for different cases,
${\cal E}_{\rm tot}$ is not the same because they will have different cosmic ray energy diffusive flux ($-\kappa\, dE_c/d\xi$).

Here we discuss solutions of the first category parameter regime.
Figure~\ref{fig:3f_diffusion} shows the cases with $\kappa=1/(3P^+_w)$ and $\gamma_g=5/3$.
All the other parameters and boundary conditions are the same as in Figure~\ref{fig:3f_nodiffusion}.
To ensure physically allowable solution, $dP_c/d\xi$ at $\xi_b$ is suitably adjusted in each case.
As a result ${\cal E}_{\rm tot}$ is slightly different in each case.
In general, the trend is the same as the cases without diffusion (see Figure~\ref{fig:3f_nodiffusion}).
If we use a constant diffusion coefficient, say $\kappa=1/3$, the resulting profiles are very similar to those of Figure~\ref{fig:3f_diffusion}.

In Figure~\ref{fig:3f_4f_no_damp}, we compare the result of a one-wave system without wave damping and without cosmic ray diffusion to
the same system but with diffusion (i.e., compare the solid lines of Figures~\ref{fig:3f_nodiffusion} \& \ref{fig:3f_diffusion}).
In the figure, solid line and dashed line correspond to one-wave system without and with diffusion.
They both have the same boundary conditions $\{U_b,P_{gb},P_{cb},P^+_{wb},\rho_b,B_b\}$ at $\xi_b$, except $dP_c/d\xi$.
As shown in the figure, in the subsonic branch (upper row), the system with diffusion has larger $U$, smaller $P_g$, larger $P_c$ and smaller $P^+_w$.
In the supersonic branch (lower row), the system with diffusion has larger $U$, smaller $P_g$, smaller $P_c$ and larger $P^+_w$ \citep[cf. e.g.,][]{Wiener_2017}.

We also give an example of a two-wave system in Figure~\ref{fig:3f_4f_no_damp} for comparison.
The two-wave system adopts Equation~(\ref{eq:epmkappatau}) (with $\nu_\pm\propto P^\pm_w$) and
with the same boundary conditions $\{U_b,P_{gb},P_{cb},P^+_{wb},\rho_b,B_b\}$ as the one-wave system,
but with an extra condition $P^-_{wb}=0.125$ for both subsonic and supersonic branches.
As shown in the figure, there are notable differences in the subsonic branch, but only tiny differences in the supersonic branch.
The insets in the figure are zoom-in versions underscoring the differences.
The upper row of Figure~\ref{fig:3f_4f_no_damp} shows that in the subsonic branch, the velocity of the two-wave system with $P^-_{wb}=0.125$ is smaller
(while the thermal pressure is larger) with respect to the one-wave system without diffusion case.
In fact, in this case as $P^-_{wb}$ increases from 0 (one-wave system) to 0.125,
the profile of $U$ ($P_g$) decreases (increases) from above (below) the case without diffusion to below (above) it.
On the other hand, in the supersonic branch, the profile of $U$ ($P_g$) increases (decreases) from above (below)
the case without diffusion onward as $P^-_{wb}$ increases from 0 to 0.125.

Figure~\ref{fig:3f_4f_NLLD}
is the same as Figure~\ref{fig:3f_4f_no_damp}
except that the wave damping mechanism nonlinear Landau damping (NLLD) is considered
($L^+_w=\delta^+\, P_w^{+\,2}\sqrt{P_g/\rho B^2}$).

\subsection{Quasi-thermal outflows}
\label{sec:quasi-thermal}

Here we discuss solutions of the second category parameter regime mentioned in Section~\ref{sec:compare_2}.
This is the regime where the waves in the system wither (through streaming instability, stochastic acceleration or other damping mechanisms).
When waves die cosmic ray will be decoupled from the thermal gas (diffusion coefficient becomes infinite, see Equation~(\ref{eq:epmkappatau})),
and the outflow becomes a thermal outflow, which we called quasi-thermal outflow.

For the purpose of illustration, we consider systems with $L^\pm_w=0$.
Suppose at large distances (subscript $a$) the wave pressures vanish and cosmic ray is decoupled from the thermal gas, then
the energy constant ${\cal E}$ of the quasi-thermal wind is
\begin{equation}\label{eq:energy_quasi_thermal}
  {\cal E}={\cal E}_{\rm tot}-\frac{\Delta_\infty}{\psi_m}F_{c\infty}
  =\frac{U^2_\infty}{2}+\frac{\gamma_g P_{g\infty}}{(\gamma_g-1)\rho_\infty}\,.
\end{equation}
As $L^\pm_w=0$, thus $P_g=P_{gb}(\rho/\rho_b)^{\gamma_g}$ (Equation~(\ref{eq:entropy})).
As the waves wither at large distances, the cosmic ray flux at large distances is related to the boundary conditions
by the wave action integral Equation~(\ref{eq:wave_action}),
\begin{align}\label{eq:CR_flux_infty}
  {\cal W}_{\rm A}
  = &\, \Delta_b\left[F_{cb}+2P^+_{wb}\frac{\left(U_b+V_{Ab}\right)^2}{V_{Ab}}-2P^-_{wb}\frac{\left(U_b-V_{Ab}\right)^2}{V_{Ab}}\right] \nonumber \\
  = &\,\Delta_\infty F_{c\infty}\,.
\end{align}
Hence
\begin{align}\label{eq:energy_quasi_thermal_b}
  {\cal E} = & \,\frac{U_b^2}{2}+\frac{U_{\rm crit}^{(\gamma_g+1)}}{(\gamma_g-1)U_b^{(\gamma_g-1)}} + \Psi_b\nonumber \\
  & -\frac{P^+_{wb}}{\rho_b}\left(1+2M_{Ab}\right)-\frac{P^-_{wb}}{\rho_b}\left(1-2M_{Ab}\right) \,,
\end{align}
where $U_{\rm crit}$ is given by Equation~(\ref{eq:critical_vel}).
By the same argument as in Appendix~\ref{sec:thermal},
the boundary conditions for physically allowable solution for the quasi-thermal outflow is constrained by
(cf. Equation~(\ref{eq:constraint}) and argument that follows)
\begin{equation}\label{eq:quasi_thermal_constraint}
  {\cal E} \ge \frac{(\gamma_g+1)U_{\rm crit}^2}{2(\gamma_g-1)}\,.
\end{equation}
If this criterion is satisfied, then the
asymptotic velocity
$U_\infty$ is given by
Equations~(\ref{eq:energy_quasi_thermal}) \& (\ref{eq:energy_quasi_thermal_b}).
The interesting fact is the criterion and the
asymptotic velocity
are independent of $P_{cb}$ (and its gradient).
The dependence of the asymptotic velocity of the quasi-thermal outflow on the boundary value at the base of the potential well is shown
in the right panel of Figure~\ref{fig:asymptotic}.
This is the same example described near the end of Section~\ref{sec:without_diffusion} on an outflow from 1 kpc above the mid-plane of our Galaxy,
except that the model for the outflow is different.
Recall that in the units for pressure, density and velocity as $10^{-12}$ erg cm$^{-3}$, $10^{-26}$ g cm$^{-3}$ and $100$ km s$^{-1}$, the values
at the base of the potential well are $\rho_b=1$, $P_{gb}=2$, $P^-_{cb}=0$, $\psi_B^\prime=B_b/\sqrt{\mu_0}=1$, $\Psi_b=-2.4$.
The dashed, long-dashed and solid lines correspond to $P^+_{wb}=0,\,0.25,\,0.5$, respectively.
As shown in the figure, the supersonic branch in the case of quasi-thermal outflow is more sensitive to $P^+_{wb}$ than in the case of one-wave system
(i.e., three-fluid model) without diffusion.

\begin{figure*}[ht]
\centering
\vskip -3.5cm
\includegraphics[width=0.9\textwidth]{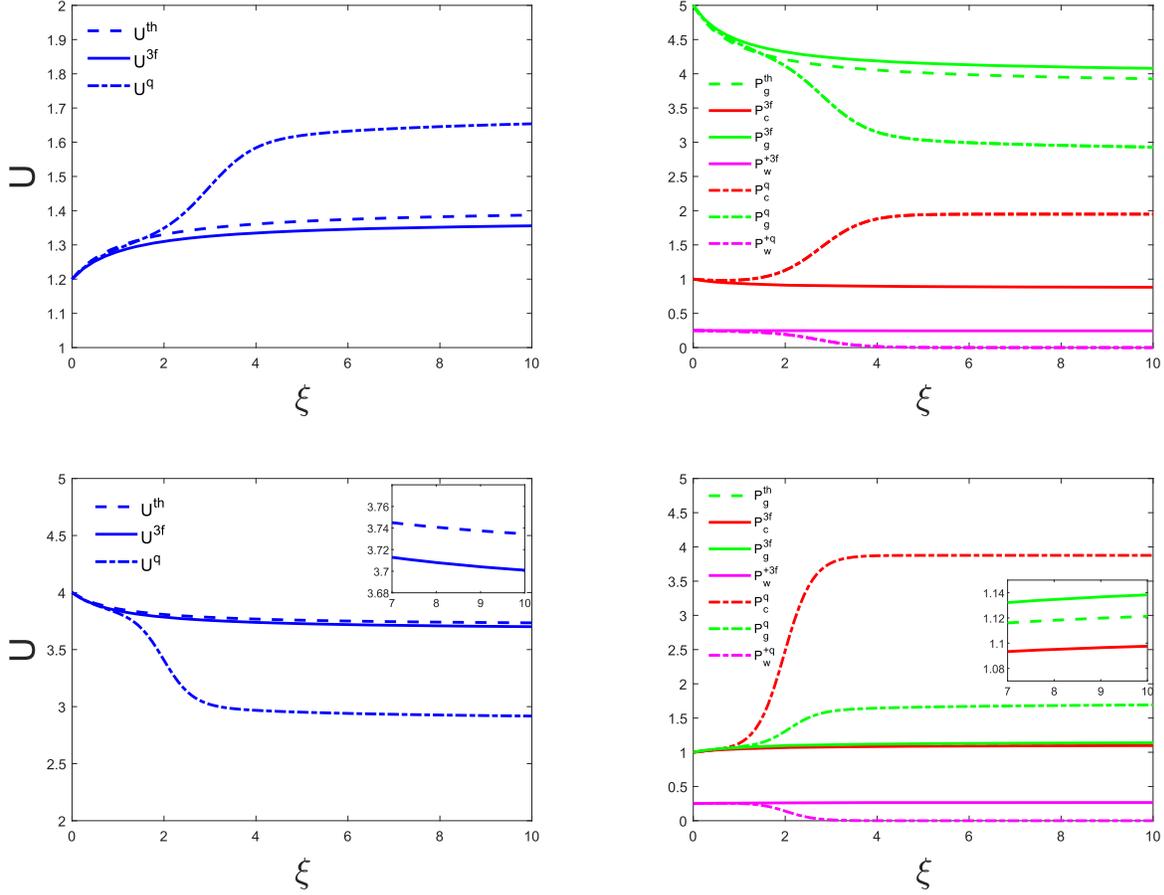}
\vskip -3.5cm
\caption{
Comparison of one-wave quasi-thermal outflow with one-wave system without diffusion in the first category parameter regime.
Both cases are without wave damping.
In this example, $\gamma_g=5/3$, $\gamma_c=4/3$, $GM=1$, $a=1$ and ${\psi_B^\prime}=1$.
{\it Upper row}: Subsonic branch with boundary conditions $\psi_m^\prime=1.2$, $U_b=1.2$, $P_{gb}=5$, $P_{cb}=1$, $P^+_{wb}=0.25$.
{\it Lower row}: Supersonic branch with boundary conditions $\psi_m^\prime=4$, $U_b=4$, $P_{gb}=1$, $P_{cb}=1$, $P^+_{wb}=0.25$.
Velocity profiles are shown on left column and pressure profiles on the right column.
Dot-dashed line and solid line denote quasi-thermal outflow and one-wave system without diffusion, respectively.
In addition, dashed lines show the profiles of velocity and thermal pressure of pure thermal outflow with $U_b=1.2$, $P_{gb}=5$ in subsonic branch,
and $U_b=4$, $P_{gb}=1$ in supersonic branch.
Note that in the supersonic branch the solid line almost overlaps the dashed line.
The insets are zoom-in versions of the figures to emphasize the slight differences between the profiles in the supersonic branch.
}
\label{fig:quasi_thermal_ND}
\end{figure*}

Figure~\ref{fig:quasi_thermal_ND} shows the profiles of an one-wave system (forward propagating wave) with $L^+_w=0$.
The upper row is an example of a subsonic flow and the lower row a supersonic flow.
The left column shows the velocity profile and the right column the pressure profiles.
Dot-dashed line is the quasi-thermal outflow (diffusive case).
For comparison we show the one-wave system without diffusion (see Section~\ref{sec:compare_1}) in solid line.
Both cases have the same boundary conditions $\{U_b,P_{gb},P_{cb},P^+_{wb},\rho_b,B_b\}$ at $\xi_b$
(the only difference is one system has diffusion and the other not).
In the figure, we also show the pure thermal outflow with the same boundary conditions $\{U_b,P_{gb},\rho_b,B_b\}$ (dashed line).

One can see in the subsonic flow, the diffusive case follows the non-diffusive case near the base.
Somewhere along the flow, a transition occurs: the diffusive case diverges from the non-diffusive case.
The wave dies, the cosmic ray pressure increases (and becomes constant),
the thermal pressure decreases, the velocity increases and the flow becomes a quasi-thermal outflow.
For the case with diffusion, the cosmic ray pressure gradient at the base must be larger
than a specific minimum value to have physically allowable solutions.
When we tune the cosmic ray pressure gradient at the base from the minimum value to some larger values
the transition region takes place closer and closer to the base,
but the asymptotic velocity and thermal pressure
($U_\infty$ and $P_{g\infty}$)
do not change.
Asymptotic cosmic ray energy flux changes according to the change in diffusive flux at the base.
Supersonic flow has similar behaviour, except that in the transition region, the thermal pressure increases while the velocity decreases.

If $L^\pm_w\ne 0$ (e.g., with INF or NLLD), the system is more complicated and cannot be analysed as nicely as the case of $L^\pm_w=0$.
In any case, parameter regions for quasi-thermal outflow exist.
However, in contrast to the simple no wave damping case,
the asymptotic velocity and thermal pressure depend on the cosmic ray pressure (and its gradient) at the base,
and of course, on the strength of damping ($\delta^\pm$).

\begin{figure*}[ht]
\centering
\vskip -4.5cm
\includegraphics[width=0.9\textwidth]{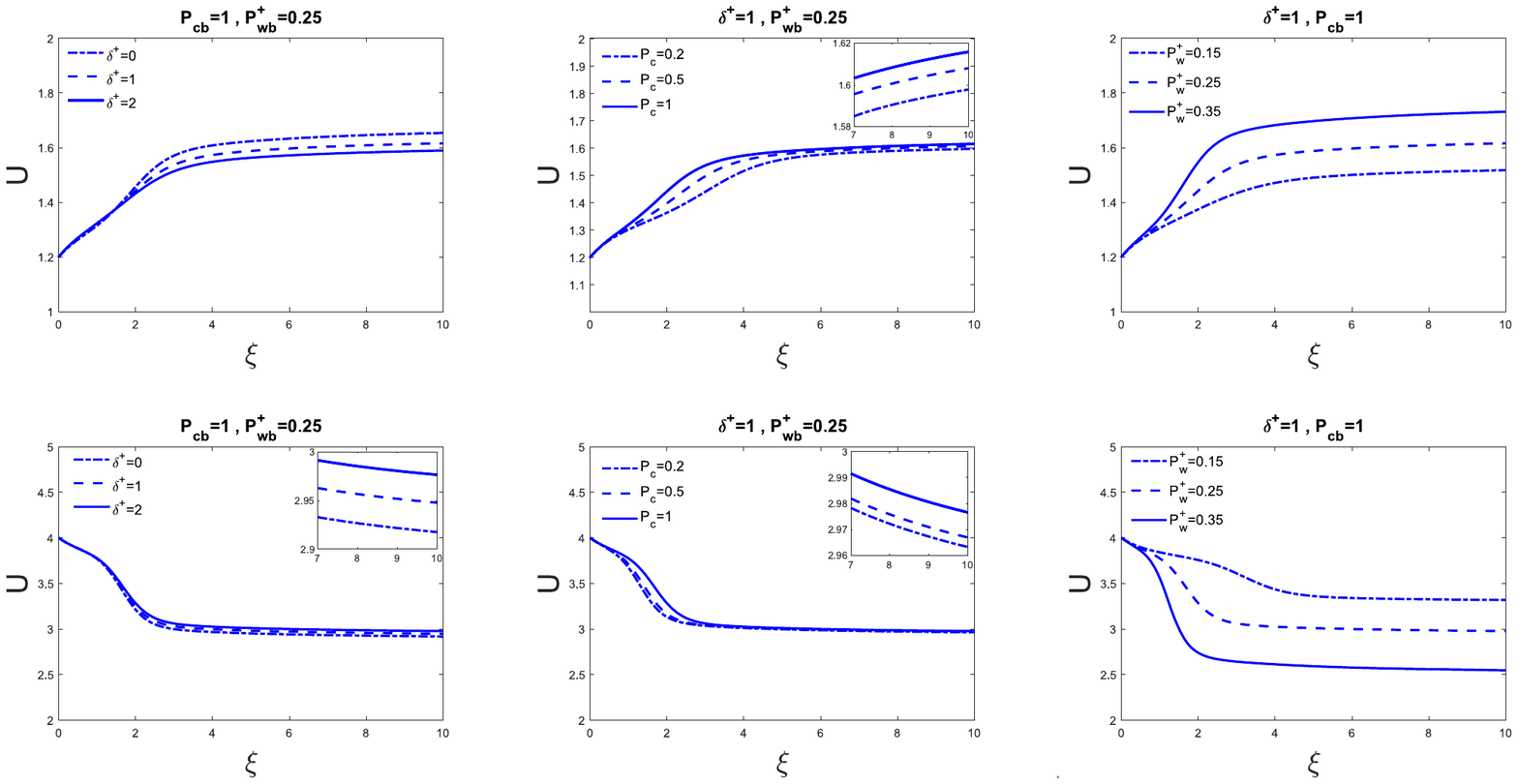}
\vskip -4.5cm
\caption{
Velocity profiles of one-wave quasi-thermal outflow with nonlinear Landau damping.
In this example, $\gamma_g=5/3$, $\gamma_c=4/3$, $GM=1$, $a=1$ and ${\psi_B^\prime}=1$.
{\it Upper row}: Subsonic branch boundary conditions $\psi_m^\prime=1.2$, $U_b=1.2$, $P_{gb}=5$, $\kappa dE_c/d\xi=0$.
{\it Lower row}: Supersonic branch boundary conditions $\psi_m^\prime=4$, $U_b=4$, $P_{gb}=1$, $\kappa dE_c/d\xi=0.4$.
Different profiles correspond to different $\delta^+$, $P_{cb}$ and $P^+_{wb}$ (see annotation in the figures for the values, or see Table\ref{tab:P_table}).
The insets are zoom-in versions of the figures to emphasize the slight differences between the velocity profiles.
}
\label{fig:quasi_thermal_WDNLLD}
\end{figure*}

An example of a one-wave quasi-thermal outflow
with NLLD is shown in
Figure~\ref{fig:quasi_thermal_WDNLLD}.
(In fact, the result is qualitatively the same with INF or constant damping rate.)
The figure illustrates how the velocity profile is affected by changing $\delta^+$, $P_{cb}$ and $P^+_{wb}$.
In the subsonic branch, as $\delta^+$ increases the velocity becomes smaller (except at regions close to the base in which the velocity
becomes larger by a small amount). As $P_{cb}$ or $P^+_{wb}$ increases, the velocity becomes larger.
In the supersonic branch, as $\delta^+$ increases the velocity becomes larger (but by a minute amount).
As $P_{cb}$ increases the velocity becomes larger (more noticeable in the transition region while only a small increase at large distances).
As $P^+_{wb}$ increases the velocity becomes quite a lot smaller.

The characteristic of the outflows described in this section is they becomes pure thermal outflows at large distances.
Now if cooling is not negligible, then the thermal outflow or wind will be significantly affected.
For instance, if we consider cooling only (and nothing else) then a one-dimensional subsonic flow will stall (see Appendix~\ref{sec:thermal}).
However, it is well known that if other effects included one-dimensional thermal wind can have interesting flow profiles, such as transonic solutions
\citep[see, e.g.,][]{Bustard_2016}.
The models described here (with wave damping or not) also possess transonic solutions.
A systematic and thorough discussion on these possible solutions warrants further investigation in the future.

\section{Summary and discussion}
\label{sec:Summary}

We study plasma outflow from a gravitational potential well with cosmic rays and self-excited Alfv\'en waves.
We explore models of different levels of complexity. The most complicated model is the four-fluid system (or called two-wave system)
which comprises thermal plasma, cosmic rays, forward and backward propagating Alfv\'en waves, with wave damping
(such as stochastic acceleration, ion-neutral friction, nonlinear Landau damping).
In principle, outflow or wind is driven by varies pressure gradients against gravity (e.g., Equation~(\ref{eq:momentum})).
Yet there are intricate interactions between the components.
In general it is difficult to analyse the four-fluid system analytically even in the so called flux-tube formation
(i.e., assume a prescribe magnetic flux tube and the flow moves along the tube).
However, for simpler systems some analytical work can be done.

We start from the simplest system we have: the three-fluid system (or one-wave system) without diffusion of cosmic rays in flux-tube formulation.
In this system we consider only forward propagating wave and assume zero cosmic ray diffusion coefficient.
The system can be analysed in exactly the same way as the classic Parker's stellar wind \citep{Parker_1958}.
The outflow velocity can be expressed (implicitly) in terms of $\xi$, the coordinate along the flux tube.
Both panels of Figure~\ref{fig:contourplot}
show the solution curves of the outflow.
For convenience we focus on one-wave system (forward wave) in constant flux tube geometry
(the right panel of Figure~\ref{fig:contourplot}).
We are interested in physically allowable solutions (a.k.a. quantities have finite value at large distances).
Figure~\ref{fig:3f_nodiffusion} shows the solution profiles of a typical set of boundary conditions
$\{U_b,P_{gb},P_{cb},P^+_{wb},\rho_b,B_b\}$ at $\xi_b$ (the base of the potential well).
For the case without wave damping, whether the set of boundary conditions can give a physically allowable solution or not
can be studied by method similar to Appendix~\ref{sec:thermal}.
Moreover, if wave damping is included, the velocity (and other quantities) may increase or decrease depending on whether the solution is subsonic or supersonic,
sub-Alfv\'enic or super-Alfv\'enic, and also some other parameters (for details, see Section~\ref{sec:compare_1}).

If cosmic ray diffusion cannot be neglected in the one-wave system, things become more interesting.
Cosmic ray diffusion is facilitated by waves, thus we take the diffusion coefficient to be inversely proportional to the wave energy density (or pressure).
Since diffusion is considered, one more condition is needed in addition to the boundary conditions mentioned above.
We simply insist on physically allowable solutions (i.e., finite value at large distances).
There are two main categories of solutions:
(1) cosmic ray accompanied outflow (solutions closely resemble the ones without diffusion), and (2) quasi-thermal outflow.

In the first category, the requirement of finite value at large distances demands a specific cosmic ray pressure gradient at the base of the potential well
(and all other values of gradient give unphysical solutions),
see Figures~\ref{fig:3f_diffusion}, \ref{fig:3f_4f_no_damp} \& \ref{fig:3f_4f_NLLD}.
The one-wave system with diffusion has larger velocity than the one without diffusion in both subsonic and supersonic branches
(but only tiny difference is observed in supersonic branch).
One-wave system is a special case of two-wave system (four-fluid system).
If we increase the amount of backward propagating waves at the base of the potential well, the velocity (thermal pressure) will decrease (increase)
accordingly in the subsonic branch, but no systematic trend in the supersonic branch (anyway the difference is minor in this branch).
A more detail description is presented in the end of Section~\ref{sec:compare_2}.

In the second category, the waves vanish at large distances and the cosmic ray is decoupled from the thermal plasma
(as the diffusion coefficient becomes very very large). The flow acts like a pure thermal outflow.
A range of cosmic ray pressure gradients at the base of the potential well can give finite value at large distances.
Basically, the solution acts like an outflow similar to the first category solutions
then undergoes a transition to a thermal outflow at large distances, see Figure~\ref{fig:quasi_thermal_ND}.

Now, if there is no extra wave damping ($L^\pm_w=0$), we can be more specific on these two categories.
If the boundary conditions $\{U_b,P_{gb},P_{cb},P^+_{wb},\rho_b,B_b\}$ satisfy/violate Equation~(\ref{eq:quasi_thermal_constraint}),
then the solutions belong to the second/first category (see Section~\ref{sec:quasi-thermal}).
Moreover, if Equation~(\ref{eq:quasi_thermal_constraint}) is satisfied, the asymptotic velocity (velocity at large distances)
is given by Equation~(\ref{eq:energy_quasi_thermal}).
The interesting fact is this
asymptotic velocity is independent of cosmic ray pressure (and its gradient) at the base of the potential well.
Nevertheless, if $L^\pm_w\ne 0$, this non-dependence of cosmic ray no longer holds.
(An example of a one-wave system with NLLD is given at the end of Section~\ref{sec:quasi-thermal}.)

To this end, we would like to address the question of the consequence of adding cosmic rays to a thermal outflow.
This depends on the system.
Suppose there is a thermal outflow with boundary conditions $\{U_b,P_{gb},\rho_b,B_b\}$.
The corresponding asymptotic velocity $U^{\rm th}_\infty$
is obtained by its energy constant (see Equation~(\ref{eq:energy_constant})).
Now cosmic ray and waves ($\{P_{cb},P^\pm_{wb}\}$) are added into the system.
Once again for convenience we consider one-wave system without wave damping ($P^-_{wb}=0$, $L^\pm_w=0$).
First, for non-diffusive case, the
asymptotic velocity $U^{3f}_\infty$
can be found by energy constant Equation~(\ref{eq:Bernoulli}).
It turns out that for most parameters (at least for $P_{cb}$ and $P^+_{wb}$ from close to zero to the order of $P_{gb}$),
$U^{3f}_\infty$ is smaller than $U^{\rm th}_\infty$
by a small amount in subsonic branch, and a very tiny amount in the supersonic branch.
If diffusion of cosmic ray cannot be ignored and the boundary conditions belong to the first category, the result is similar to the non-diffusive case.
However, it is more interesting if the boundary condition belong to the second category (the quasi-thermal case).
The asymptotic velocity $U^{\rm q}_\infty$
is given by the energy constant Equations~(\ref{eq:energy_quasi_thermal}) \& (\ref{eq:energy_quasi_thermal_b}).
In this case,
$U^{\rm q}_\infty>U^{\rm th}_\infty$ in the subsonic branch and $U^{\rm q}_\infty<U^{\rm th}_\infty$ in the supersonic branch.
Figure~\ref{fig:quasi_thermal_ND} shows an example of one of these cases.
We note that an exposition of this question from other perspective is given at Appendix~\ref{sec:thermal}.

\section*{Acknowledgments}

BR and CMK are supported in part by the Taiwan Ministry of Science and Technology grants
MOST 105-2112-M-008-011-MY3, MOST 108-2112-M-008-006 and MOST 109-2112-M-008-005.
DOC is supported by the grant RFBR 18-02-00075 and by foundation for the advancement of theoretical physics ``BASIS''.

\appendix

\section{Thermal outflows}
\label{sec:thermal}

In this appendix we revisit the classical thermal outflow or wind in a potential well.
We examine the characteristic of the solution from a given set of boundary conditions at the base of the potential well.
Specifically, we study the constraint on the velocity at the base in order to have physical outflow solution when the thermal pressure at the base is given.

For simplicity, we consider polytropic gas in one-dimensional geometry (constant area flow tube, i.e., $\Delta$ is a constant). The wind equation is
\begin{equation}\label{eq:wind_thermal}
  \left(1- M^{-2}\right)U\frac{dU}{d\xi} = -\frac{d\Psi}{d\xi}\,,
\end{equation}
where $M=U/a_g$, $a_g^2=\gamma_g P_g/\rho=\gamma_g P_g U/\psi_m^\prime$, $\psi_m^\prime=\rho U$ is the mass flux which is a constant in one-dimensional flow.
Note that the mass flow rate $\psi_m=\rho U\Delta=\psi_m^\prime\Delta$.
$\Psi$ is the potential well, which is a monotonically increasing function of $\xi$ from the base of the potential well $\xi_b$ to $\xi\rightarrow\infty$.
We set $\Psi=0$ as $\xi\rightarrow\infty$, thus $\Psi\le 0$.
Suppose at the base of the potential well, $U=U_b$, $\rho=\rho_b$, $P_g=P_{gb}$ (and $\psi_m^\prime=\rho_b U_b$),
and for convenience we call these ``boundary conditions''.

\begin{figure*}[ht]
\centering
\includegraphics[width=0.8\textwidth]{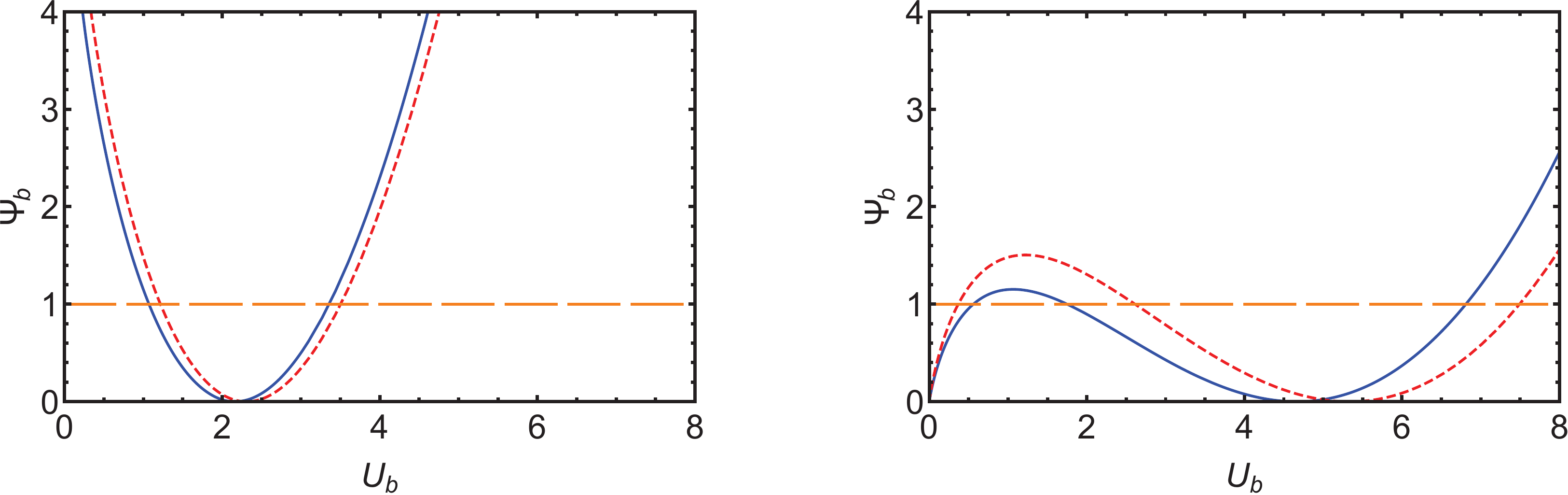}
\caption{Constraint on velocity $U_b$ at the base of the potential well $\Psi_b$. The long dashed orange line represent $-\Psi_b$.
{\it Left}: $\{P_{gb},\rho_b\}=\{2.8,1\}$ and $\{P_{gb},\rho_b\}=\{3.2,1\}$ for the blue curve and dashed red curve, respectively.
{\it Right}: $\{P_{gb},\psi_m^\prime\}=\{2.8,1\}$ and $\{P_{gb},\psi_m^\prime\}=\{3.2,1\}$ for the blue curve and dashed red curve, respectively.
$\gamma_g=5/3$ for both cases.
The intersection of the curve and the long dashed orange line is the value of $U_b$ that gives rise to a transonic solution.}
\label{fig:constraint}
\end{figure*}

The sonic point of the system is located at $\xi\rightarrow\infty$ or $\Psi=0$, and the critical velocity is given by $U_{\rm crit}^2=a_g^2$, or
\begin{equation}\label{eq:critical_vel}
  U_{\rm crit}^{(\gamma_g+1)} = \frac{\gamma_g P_{gb}U_b^{(\gamma_g-1)}}{\rho_b} = \frac{\gamma_g P_{gb}U_b^{\gamma_g}}{\psi_m^\prime}\,.
\end{equation}
The solution to the wind equation can be expressed in terms of an integral, energy constant
\begin{equation}\label{eq:energy_constant}
  \frac{U^2}{2}+\frac{U_{\rm crit}^{(\gamma_g+1)}}{(\gamma_g-1)U^{(\gamma_g-1)}}+\Psi = {\cal E}\,,
\end{equation}
and for polytropic gas $\gamma_g>1$.
The correspond energy constant for transonic solution is
\begin{equation}\label{eq:transonic_energy}
  {\cal E}_{\rm crit}=\frac{(\gamma_g+1)U_{\rm crit}^2}{2(\gamma_g-1)}\,.
\end{equation}
Equations~(\ref{eq:energy_constant}) \& (\ref{eq:transonic_energy}) give the constraint on the boundary conditions for possible transonic solution,
\begin{equation}\label{eq:constraint}
  \frac{U_b^2}{2}+\frac{U_{\rm crit}^{(\gamma_g+1)}}{(\gamma_g-1)U_b^{(\gamma_g-1)}}-\frac{(\gamma_g+1)U_{\rm crit}^2}{2(\gamma_g-1)} = -\Psi_b\,.
\end{equation}
In Figure~\ref{fig:constraint} the blue and red (dashed) curves are the plots of the LHS of Equation~(\ref{eq:constraint}) as a function of $U_b$.
In the left panel, $P_{gb}$ and $\rho_b$ are given while in the right panel $P_{gb}$ and $\psi_m^\prime$ are given.
We set $\Psi_b=-1$ as the base of the potential well and is the long dashed orange line in the figure.
The intersections of the curves and the long dashed orange line are the values of the velocity at the base ($U_b$) that give rise to transonic solutions.
For physically allowable solutions (i.e., solutions that extend to large distances or $\Psi=0$), ${\cal E}\ge {\cal E}_{\rm crit}$.
Thus, the ranges of $U_b$ in which the curve is above the long dashed orange line are those boundary conditions correspond to physically allowable solutions.
The range with smaller $U_b$ is subsonic and the range with larger $U_b$ is supersonic.

In the figure, $P_{gb}$ of the red dashed curve is larger than that of the blue curve.
As shown in the figure, the range of the subsonic/supersonic branch of the red dashed curve is larger/smaller than the blue curve.
Thus for subsonic/supersonic solutions close to transonic solution, decreasing/increasing $P_{gb}$ (but keeping $U_b$ and $\rho_b$ fixed)
nudges the solution to unphysical regime.

In the case of given $P_{gb}$ and $\psi_m^\prime$ (right panel of the figure), there is only supersonic branch if $P_{gb}$ is small enough,
or $\psi_m^\prime$ is large enough or the potential well is deep enough (i.e., more negative $\Psi_b$).
In the case of given $P_{gb}$ and $\rho_b$ (left panel of the figure), there is always a subsonic branch and a supersonic branch
if $\gamma_g P_{gb}/(\gamma_g-1)\rho_b+\Psi_b\ge 0$, otherwise there is only supersonic branch.

Similar analysis can be applied to one-wave system without cosmic ray diffusion described in the main text.
As expected, similar conclusion is obtained.
We like to know the consequence of adding cosmic rays to a thermal wind. This depends on how we frame the question.
\begin{itemize}
\item {Prescribed $P_{gb}$, $\rho_b$, $U_b$ (i.e., the mass flux $\psi_m^\prime$ is fixed also):}

If the thermal wind is in the supersonic region and is close to a transonic solution,
then adding (a small amount of) cosmic ray may nudge it towards the unphysical regime.
On the other hand, if the thermal wind is in the subsonic region but within the unphysical regime, then adding cosmic ray may push the flow
to enter the physically allowable regime.

\item {Prescribed $P_{gb}$, $\rho_b$, but $U_b$ is allowed to change (i.e., the mass flux $\psi_m^\prime$ can be adjusted):}

Suppose the thermal wind is at a transonic solution. Now add (a small amount of) cosmic ray and then seek a new transonic solution.
The new $U_b$ (and the profile as well) will be higher.
\end{itemize}

Suppose cooling is significant and cannot be neglected in the flow, then Equation~(\ref{eq:wind_thermal}) should be replaced by
\begin{equation}\label{eq:entropy_cooling}
  \frac{d}{d\xi}\left(P_g \rho^{-\gamma_g}\right) =  \frac{(\gamma_g-1)\rho^{-\gamma_g}}{U}\, \Gamma\,,
\end{equation}
\begin{equation}\label{eq:wind_cooling}
  \left(1- M^{-2}\right)U\frac{dU}{d\xi} = -\frac{d\Psi}{d\xi}\, - \frac{(\gamma_g-1)}{\rho U}\, \Gamma\,,
\end{equation}
where $\Gamma<0$ for cooling. For example, for radiative cooling $\Gamma=-\rho^2\Lambda(T)$, where $T\propto P_g/\rho$ is temperature.
When cooling (and heating for that matter) is not negligible, the integral energy constant ${\cal E}$ no longer exists.
However, if gravity can be neglected (e.g., at a large distances from the base of the potential well), then the total momentum
is conserved in one-dimensional flow,
\begin{equation}\label{eq:momentum_integral}
  \rho U^2+P_g=\psi_m^\prime U+P_g=G=\psi_m^\prime G^\prime\,.
\end{equation}
The Mach number becomes
\begin{equation}\label{eq:Mach_number}
  M^2=\frac{U^2}{a_g^2}=\frac{\rho U^2}{\gamma_g P_g}=\frac{U}{\gamma_g(G^\prime-U)}\,,
\end{equation}
and $M$ is a monotonically increasing function of $U$ for $0\le U<G^\prime$.
Moreover, $T\propto U(G^\prime-U)$, i.e., $T\rightarrow 0$ as $U\rightarrow 0$ or $G^\prime$.

For cooling, $\Gamma<0$ and the right hand side of Equation~(\ref{eq:wind_cooling}) is positive if gravity is negligible.
Therefore, in the subsonic regime, $dU/d\xi<0$. The velocity decreases monotonically along the flow and the flow becomes more and more subsonic
(see Equation~(\ref{eq:Mach_number})). Eventually, $U\rightarrow 0$ and $\rho\rightarrow\infty$ (and $T\rightarrow 0$).
The flow will stall, unless $\Gamma$ vanishes before $U=0$.

On the other hand, in the supersonic regime, $dU/d\xi>0$. The velocity increases monotonically along the flow and the flow becomes more and more supersonic.
Eventually, $U\rightarrow G^\prime$ and $P_g\rightarrow 0$ (and $T\rightarrow 0$).
The flow will be unphysical as the thermal pressure will become negative (see Equation~(\ref{eq:entropy_cooling})),
unless $\Gamma$ vanishes before $U=G^\prime$.

Now let us put gravity back into consideration.
If the gravity strong enough, then it is possible that the right hand side of Equation~(\ref{eq:wind_cooling}) is negative near the base
and positive at large distances, i.e., it is zero at some finite distance.
Thus, it is conceivable that a sonic point exists at finite distance (the right hand side of Equation~(\ref{eq:wind_cooling}) vanishes together with $M=1$).
Therefore, transonic solutions are possible.


\begin{thebibliography}{}

\bibitem[Barros et al.(2016)]{Barros_2016}
Barros, D. A., L\'epine, J. R. D., \& Dias, W. S. 2016, \aap, 593, A108

\bibitem[Bondi(1952)]{Bondi_1952}
Bondi, H. 1952, MNRAS, 112, 195

\bibitem[Breitschwerdt et al.(1991)]{Breitschwerdt_1991}
Breitschwerdt, D., McKenzie, J. F., \& V\"olk, H. J. 1991, \aap, 245, 79

\bibitem[Breitschwerdt et al.(1993)]{Breitschwerdt_1993}
Breitschwerdt, D., McKenzie, J. F., \& V\"olk, H. J. 1993, \aap, 269, 54

\bibitem[Bustard et al.(2016)]{Bustard_2016}
Bustard, C.,  Zweibel, E. G., \& D’Onghia, E. 2016, \apj, 29, 819

\bibitem[Cox(2005)]{Cox_2005}
Cox, D. P. 2005, \araa, 43, 1

\bibitem[Dorfi \& Breitschwerdt(2012)]{Dorfi_2012}
Dorfi, E. A., \&  Breitschwerdt, D. 2012, \aap, 540, A77

\bibitem[Dorfi et al.(2019)]{Dorfi_2019}
Dorfi, E. A., Steiner, D., Ragossnig, F., \&  Breitschwerdt, D. 2019, \aap, 630, A107

\bibitem[Drury \& V\"olk(1981)]{Drury_1981}
Drury, L. O'C., \& V\"olk, H. J. 1981, \apj, 248, 344.


\bibitem[Everett et al.(2008)]{Everett_2008}
Everett, J. E., Zweibel, E. G., Benjamin, R. A., et al. 2008, \apj, 674, 258

\bibitem[Farber et al.(2018)]{Farber_2018}
Farber, R., Ruszkowski, M., Yang, H.-Y. K., \& Zweibel, E. G. 2018, \apj, 856, 112

\bibitem[Ferri\`ere (2001)]{Ferrire_2001}
Ferri\`ere, K. M. 2001, \rmp, 73, 4

\bibitem[Ghosh \& Ptuskin(1983)]{Ghosh_1983}
Ghosh, A., \& Ptuskin V. S. 1983, \apss, 92, 37

\bibitem[Ginzburg \& Ptuskin(1976)]{Ginzburg_1976}
Ginzburg, V.~L., \& Ptuskin, V.~S. 1976, \rmp, 48, 2

\bibitem[Girichidis et al.(2016)]{Girichidis_2016}
Girichidis, P., Naab, T., Walch, S., et al. 2016, \apjl, 816, L19

\bibitem[Hanasz \& Lesch(2000)]{Hanasz_2000}
Hanasz, M., \& Lesch, H. 2000, \apj, 543, 235

\bibitem[Hanasz \& Lesch(2003)]{Hanasz_2003}
Hanasz, M., \& Lesch, H. 2003, \aap, 412, 331

\bibitem[Heintz \& Zweibel(2018)]{Heintz_2018}
Heintz, E., \& Zweibel, E. G. 2018, \apj, 860, 97

\bibitem[Heintz et al.(2020)]{Heintz_2020}
Heintz, E., Bustard, C., \& Zweibel, E. G. 2020, \apj, 891, 157

\bibitem[Holguin et al.(2019)]{Holguin_2019}
Holguin, F., Ruszkowski, M., Lazarian, A., Farber, R., \& Yang, H.-Y. K. 2019, \mnras, 490, 1271

\bibitem[Ipavich(1975)]{Ipavich_1975}
Ipavich, F. M. 1975, \apj, 196, 107

\bibitem[Ko(1991a)]{Ko_1991a}
Ko, C. M. 1991a, \aap, 242, 85

\bibitem[Ko(1991b)]{Ko_1991b}
Ko, C. M. 1991b, \aap, 251, 713

\bibitem[Ko(1992)]{Ko_1992}
Ko, C. M. 1992, \aap, 259, 377

\bibitem[Ko(2001)]{Ko_2001}
Ko, C. M. 2001, J. Plasma Phys., 65, 305

\bibitem[Ko \& Webb(1987)]{Ko_1987}
Ko, C. M., \& Webb, G. M. 1987, \apj, 323, 657

\bibitem[Ko et al.(1988)]{Ko_1988}
Ko, C. M., Jokipii, J. R., \& Webb, G. M. 1988, \apj, 326, 761

\bibitem[Ko et al.(1991)]{Koetal_1991}
Ko, C. M., Dougherty, M.~K., \&  McKenzie, J. F. 1991, \aap, 241, 62

\bibitem[Ko et al.(1997)]{Ko_1997}
Ko, C. M., Chan, K.W., \& Webb, G. M. 1997, J. Plasma Phys., 57, 677

\bibitem[Ko \& Lo(2009)]{Ko_2009}
Ko, C. M., \& Lo, Y. Y. 2009, \apj, 691, 2

\bibitem[Kulsrud(2005)]{Kulsrud_2005}
Kulsrud, R. M. 2005, Plasma Physics for Astrophysics (Princeton, NJ: Princeton Univ. Press)


\bibitem[Kuwabara et al.(2004)]{Kuwabara_2004}
Kuwabara, T., Nakamura, K., \& Ko, C. M. 2004, \apj, 607, 828

\bibitem[Kuwabara \& Ko(2006)]{Kuwabara_2006}
Kuwabara, T., \& Ko, C. M. 2006, \apj, 636

\bibitem[Kuwabara \& Ko(2015)]{Kuwabara_2015}
Kuwabara, T., \& Ko, C. M. 2015, \apj, 798, 79

\bibitem[Kuznetsov \& Ptuskin(1983)]{Kuznetsov_1983}
Kuznetsov, V. D., \& Ptuskin, V. S. 1983, \apss, 94, 5

\bibitem[Lo et al.(2011)]{Lo_2011}
Lo, Y. Y, Ko, C. M., \& Wang, C. Y. 2011, Computer Physics Communications, 182, 177

\bibitem[Mao \& Ostriker(2018)]{Mao_2018}
Mao, S. A., \& Ostriker, E. C. 2018, \apj, 854, 89

\bibitem[McKenzie \& V\"olk(1982)]{McKenzie_1982}
McKenzie, J. F., \& V\"olk, H. J. 1982, \aap, 116, 191

\bibitem[Parker(1958)]{Parker_1958}
Parker, E. N. 1958, \apj, 128, 664

\bibitem[Parker(1963)]{Parker_1963}
Parker, E. N. 1963, Interplanetary Dynamical Processes (New York: Interscience Publ.)

\bibitem[Parker(1966)]{Parker_1966}
Parker, E. N. 1966, \apj, 145, 811

\bibitem[Parker(1969)]{Parker_1969}
Parker, E. N. 1969, \ssr, 9, 3

\bibitem[Recchia et al.(2016)]{Recchia_2016}
Recchia S., Blasi, P., \& Morlino G., 2016, \mnras, 462, 4227

\bibitem[Ruszkowski et al.(2017)]{Ruszkowski_2017}
Ruszkowski, M., Yang, H.-Y. K., \& Zweibel, E. 2017, \apj, 834, 208

\bibitem[Ryu et al.(2003)]{Ryu_2003}
Ryu, D., Jongsoo, K., Seung S. H., \& Jones, T. W. 2003, \apj, 589, 1

\bibitem[Skilling(1975)]{Skilling}
Skilling, J. 1975, \mnras, 172, 557



\bibitem[Wiener et al.(2017)]{Wiener_2017}
Wiener, J., Pfrommer, C., \& Oh, S. P. 2017, \mnras, 467, 646

\bibitem[Yang et al.(2012)]{Yang_2012}
Yang, H.-Y. K., Ruszkowski, M.,  Ricker, P. M., Zweibel, E., \& Lee, D. 2012, \apj, 761, 2

\bibitem[Yu et al.(2020)]{Yu_2020}
Yu, B. P. B., Owen, E. R.,  Wu, K., \& Ferreras, I. 2020, \mnras, 492, 3179

\bibitem[Zhang(2018)]{Zhang_2018}
Zhang, D. 2018, Galaxies, 6, 114

\bibitem[Zirakashvili et al.(1996)]{Zirakashvili_1996}
Zirakashvili, V. N.,  Breitschwerdt, D., Ptuskin, V. S., \&  V\"olk, H. J. 1996, \aap, 311, 113

\bibitem[Zweibel(2017)]{Zweibel_2017}
Zweibel, E. G. 2017, PhPl, 24, 5402Z

\end{thebibliography}
\bibliographystyle{apj}

\end{document}